\documentclass[%
 reprint,
superscriptaddress,
 amsmath,amssymb,
 aps,
pre,
]{revtex4-2}

\usepackage{graphicx}
\usepackage{dcolumn}

\usepackage{bbold}

\usepackage{tikz}

\tikzset{every picture/.style={line width=0.75pt}} 

\graphicspath{ {./images/} }

\usepackage{mathtools}

\usepackage{enumitem}

\usepackage{breqn}

\usepackage{amsmath}

\usepackage{comment}

\usepackage{lipsum} 

\usepackage{graphicx,capt-of}

\usepackage{ulem}

\usepackage{duckuments}

\DeclareMathAlphabet\mathbfcal{OMS}{cmsy}{b}{n}

\DeclareMathOperator{\arccosh}{arcCosh}

\usepackage[hidelinks]{hyperref}

\newcommand{\sref}[1]{Section.~\ref{#1}}
\newcommand{\fref}[1]{Fig.~\ref{#1}}

\begin{document}

\preprint{APS/123-QED}

\title{Dynamical phase transitions in certain non-ergodic stochastic processes}

\author{Yogeesh Reddy Yerrababu}
 \email{yogeesh.yerrababu@gmail.com}
 \let\comma,
\affiliation{University of Italian Switzerland\comma \ 6900 Lugano\comma \ Switzerland}%
\affiliation{Department of Theoretical Physics\comma \ Tata Institute of Fundamental Research\comma \ Mumbai 400005\comma \ India}%
  \author{Satya N. Majumdar}%
 \email{satyanarayan.majumdar@cnrs.fr}
 \let\comma,
\affiliation{LPTMS\comma\ CNRS\comma \ Université Paris-Saclay\comma \ 91405 Orsay\comma \ France}%
  \author{Benjamin Guiselin}%
\email{benjamin.guiselin@umontpellier.fr}
 \let\comma,
\affiliation{Laboratoire Charles Coulomb (L2C)\comma\ Universit\'e de Montpellier\comma\ CNRS\comma\ Montpellier\comma\ France.}%
\author{Tridib Sadhu}%
 \email{tridib@theory.tifr.res.in}
  \let\comma,
\affiliation{Department of Theoretical Physics\comma \ Tata Institute of Fundamental Research\comma \ Mumbai 400005\comma \ India}%


\begin{abstract}
We present a class of stochastic processes in which the large deviation functions of time-integrated observables exhibit singularities that relate to dynamical phase transitions of trajectories. These illustrative examples include Brownian motion with a death rate or in the presence of an absorbing wall, for which we consider a set of empirical observables such as the net displacement, local time, residence time, and area under the trajectory. Using a backward Fokker-Planck approach, we derive the large deviation functions of these observables, and demonstrate how singularities emerge from a competition between survival and diffusion. Furthermore, we analyse this scenario using an alternative approach with tilted operators, showing that at the singular point, the effective dynamics undergoes an abrupt transition. Extending this approach, we show that similar transitions may generically arise in Markov chains with transient states. This scenario is robust and generalizable for non-Markovian dynamics and for many-body systems,
potentially leading to multiple dynamical phase transitions. We have confirmed most of our findings on the singular large-deviation function using rare-event simulation techniques.

\end{abstract}


\maketitle

\section{Introduction}
Dynamical phase transitions (DPTs) are singular changes in the distribution of dynamical observables, manifesting as singularities in their large deviation functions. These transitions are observed in systems with many degrees of freedom, such as driven diffusive systems \cite{baek2015singularities,baek2017dynamical,bertini2010lagrangian,bunin2013cusp,aminov2014singularities,shpielberg2017numerical,kumar2011symmetry}, lattice gas models \cite{bodineau2005distribution,bertini2005current,bertini2006non}, kinetically constrained models of glasses \cite{garrahan2007dynamical,garrahan2009first}, interface models \cite{le2016large,janas2016dynamical}, active matter \cite{cagnetta2017large,nemoto2019optimizing}, random graphs \cite{coghi2019large}, and others \cite{szavits2014constraint,jack2013large,bodineau2007cumulants,prolhac2009cumulants,hurtado2011spontaneous,shpielberg2018universality,bodineau2012finite}. In systems with a few degrees of freedom, DPTs were observed in the weak-noise limit of Langevin dynamics \cite{baek2015singularities,speck2012large,nyawo2016large}. In recent years, there have been reports of simple models having DPTs without requiring macroscopic or low noise limits, such as in the context of reset processes \cite{majumdar2015dynamical,pal2016diffusion,harris2017phase,majumdar2015random,santra2022effect,smith2022condensation,biroli2023extreme,gupta2019stochastic,gupta2020stochastic,di2023current}, run-and-tumble dynamics \cite{mori2021condensation,mori2021first,gradenigo2019first,mallmin2019comparison,santra2020run,proesmans2020phase,banerjee2020current,mukherjee2024large}, constrained Brownian motion\cite{smith2019geometrical,meerson2019geometrical}, drifted Brownian motion \cite{nyawo2017minimal,nyawo2018dynamical,majumdar2020statistics}, driven random walker  \cite{mallmin2019comparison,whitelam2021varied,gingrich2014heterogeneity}, vicious Brownian walkers \cite{mukherjee2023dynamical}, active Brownian particles \cite{majumdar2020toward}, Brownian motion with dry friction \cite{szavits2015inequivalence}, Brownian motion with certain special observables \cite{kanazawa2024dynamical,kanazawa2024universality} or in higher dimensions \cite{kanazawa2024universality}.

In this paper, we argue that DPTs could generically arise in certain non-ergodic stochastic processes without the need for weak-noise or macroscopic limit.
This scenario provides a unifying narrative for the DPTs reported earlier in several seemingly unrelated models \cite{majumdar2015dynamical,pal2016diffusion,gupta2019stochastic,mallmin2019comparison,nyawo2018dynamical,nyawo2017minimal,whitelam2021varied,szavits2015inequivalence,mukherjee2023dynamical,mukherjee2024large}. Understanding the mechanism helps us construct interesting phase behaviours for many-body systems and their extensions in non-Markovian processes.

We present the scenario for illustrative examples of one-dimensional Brownian motion, $X_t$, in which non-ergodicity is introduced by a leaking probability in the dynamics, caused, for example, by an absorption site. We consider several different empirical observables for the process $X_t$, which are of the form \cite{jack2010large, jack2015effective,chetrite2013nonequilibrium,chetrite2015nonequilibrium,derrida2019large}
\begin{equation}
    Q=\int_0^T dt\; U(X_t)+\int_0^TdX_t \; V(X_t),
    \label{eqn:residencetime}
\end{equation}
where $U$ and $V$ are functions of $X_t$. The second integral in \eqref{eqn:residencetime} is interpreted as a stochastic integral with a Stratonovich discretisation. 

The dynamical observable $Q$ is a functional of the Brownian trajectory and a fluctuating quantity that depends on a particular realization of the process. The probability of $Q$ can be systematically analyzed  using a well-established framework for Brownian functionals \cite{majumdar2005brownian}. In our explicit examples, we consider four different observables for the Brownian motion: (a) net displacement ($U=0$, $V=1$), (b) local time at position $a$ ($U(x)=\delta(x-a)$, $V=0$), (c) residence time in an interval ($U(x)=1$ for $a<x<b$ and zero outside, $V=0$), and (d) area under the  Brownian trajectory ($U=x$ and $V(x)=0$).

We explore these observables for two specific Brownian dynamics. In the first example, a mortal Brownian particle has a probability of dying rendering it immobile. In the second example, the Brownian motion is constrained to move between a fixed absorbing wall and a reflecting wall. For both examples, we show that the probability $P_T(Q)$ of $Q$ typically (with exceptions discussed in the text) takes a large deviation form for large $T$,
\begin{equation}
    P_T(Q=qT)\sim e^{-T\phi(q)},
    \label{eqn:largedeviation}
\end{equation}
where $\phi(q)$ is the large deviations function (ldf). For the four observables defined above, we determine $\phi(q)$ using the framework in \cite{majumdar2005brownian} and show that for all four observables, $\phi(q)$ is singular at certain values of $q$.

These singularities are linked to the dynamical phase transitions, arising from the interplay between trajectories that persist without being dead or absorbed throughout the entire duration $T$ and trajectories that do not survive. Depending on the value of $Q$, one of these two kinds of trajectories is most probable for large $T$, and the singular point in $\phi(q)$ marks the transition between these two types of evolutions. This represents a path-space generalization of conventional equilibrium phase transitions, where singularity in thermodynamic free energy indicates transitions in static observables. 

An insightful approach to quantify the differences in the nature of trajectories is in terms of an effective dynamics \cite{chetrite2013nonequilibrium,majumdar2015effective,jack2015effective,derrida2019large,de2021generating,de2021generating2,Grela2021}. Employing a theory of constrained dynamics \cite{jack2010large,chetrite2013nonequilibrium,chetrite2015nonequilibrium,derrida2019large}, we show that in our examples, the effective dynamics attaining a value of $Q$ for large $T$ is described by a Langevin process, whose nature changes across the DPT.

The theory of constrained dynamics \cite{jack2010large, jack2015effective,chetrite2013nonequilibrium,chetrite2015nonequilibrium,derrida2019large} draws parallels between the mechanism of DPTs and equilibrium phase transitions. In equilibrium lattice models, the singularity of the free energy at a phase transition is associated with the crossing of leading eigenvalues of a transfer matrix \cite{Goldenfeld,Dhar2011,cuesta2004general}. The singularity of $\phi(q)$ in DPTs is related to a similar crossing of two largest eigenvalues of a tilted operator. Typically, such a crossing of eigenvalues is forbidden by the Perron-Frobenius-Jentz theorem \cite{cuesta2004general}, unless the tilted operator falls outside the domain of the theorem. This indeed occurs for reducible tilted operators in non-ergodic stochastic processes like those in our examples. This offers a unifying perspective on the origin of DPTs in non-ergodic stochastic processes.

We illustrate the generality of this mechanism in Markov chains with transient state spaces where eigenvalue crossing results in DPTs of observables that are discrete analogs of \eqref{eqn:residencetime}. The mechanism extends to many-body systems and non-Markovian processes. Our work asserts earlier observations \cite{nyawo2017minimal,nyawo2018dynamical,mallmin2019comparison,mukherjee2023dynamical,whitelam2021varied,szavits2015inequivalence,gingrich2014heterogeneity} of DPTs in non-ergodic dynamics. All our results about  singular large-deviation functions have been validated using rare-event simulation techniques \cite{hartmann2015big,Burenev2024_importance_sampling}. Similar analysis was done earlier in a  related problems \cite{hartmann2023distribution}.

The paper is organized in the following order. After a brief introduction about relevant concepts of large deviations in Sec \ref{sec:ldf}, we discuss DPTs for mortal Brownian motion in Sec \ref{sec:stickybrownian} and for Brownian motion in presence of an absorbing boundary in Sec \ref{sec:BM abs}. These results obtained using the backward Fokker-Planck approach \cite{majumdar2005brownian} are then reproduced using an alternative approach of tilted operators in Sec \ref{sec:tiltedintro}. An extension of the DPTs for Markov chains is discussed in Sec \ref{sec:markovchains} with a general scenario presented in Sec \ref{sec:generalidea}. In Sec \ref{sec:multipledpts}, we present examples of many-particle systems with a sequence of DPTs. Examples of DPTs in non-Markov processes is discussed in Sec \ref{sec:nonmarkovextend}, with applications in fractional Brownian motion and in a model of active matter. We conclude in Sec \ref{sec:conclusions} with a discussion about open directions. Details of the importance-sampling methods for numerical computation of the large-deviation functions are presented in the Appendix.

\section{Large deviation theory \label{sec:ldf}}
The large deviation asymptotic in \eqref{eqn:largedeviation} precisely means
\begin{equation}
    \lim_{T\to\infty} \frac{1}{T}\ln P_T(Q=qT)=- \phi(q).
\label{eqn:ratefunction}
\end{equation}
This is equivalent to an asymptotic of the generating function
\begin{equation}
    \langle e^{p Q}\rangle\sim e^{T \mu(p)}
\label{eqn:scgf}
\end{equation}
for large $T$, where $\mu(p)$ is the scaled cumulant generating function (scgf). The scgf relates to the large deviation function $\phi(q)$ by a Legendre-Fenchel transformation
\begin{equation}
    \mu(p)=\max_{q}\{pq-\phi(q)\}.
\label{eqn:legendre}
\end{equation}
For an introduction to the large deviation theory see \cite{touchette2009large,majumdar2017large}.

\section{Mortal Brownian motion}\label{sec:stickybrownian}
The simplest continuous model we consider is the motion of a one-dimensional Brownian particle with a probability rate $\alpha$ of dying, after which the particle remains immobile forever. In a small time interval $dt$, the position $x_t$ of an alive particle evolves via the Langevin equation
\begin{equation}
    x_{t+dt}= \begin{dcases} x_t & \textrm{with prob }\alpha dt \ \textrm{of death} \\
    x_t+ \eta_t dt& \textrm{with prob} \ 1-\alpha dt, \end{dcases}\label{eq:sticky langevin}
\end{equation}
where $\eta_t$ is a Gaussian white noise with zero mean and the correlator
$\langle \eta_t\eta_{t'}\rangle= 2 D \delta(t-t')$. For simplicity, we henceforth set $D=1$ for the rest of the paper. In the above equation, once the particle dies, $x_{t+dt}=x_t$ for the rest of time. Similar diffusion processes with mortality have been studied in related contexts \cite{mazzolo2022conditioning,berman1996distributions,holcman2005survival,frydman2000gussian,chen2022most}.

\begin{subequations}The dynamics can also be described by the Fokker-Planck equation for the probability $P_t^{(a)}(x)$ for the particle to start at the origin and remain alive until time $t$, at which point it is at position $x$,
\begin{equation}
    \partial_t P^{(a)}_t(x)=\partial_{x}^2 P^{(a)}_t(x) - \alpha P^{(a)}_t(x)
\end{equation}
with the initial condition $P_0^{(a)}(x)=\delta(x)$.
The probability for the particle to be dead at position $x$ at time $t$ follows
\begin{equation}
    \partial_t P^{(d)}_t(x)= \alpha P^{(a)}_t(x)
\end{equation}\label{eqn:sticky pos eqns}\end{subequations}
with the initial condition $P_0^{(d)}(x)=0$. 
The probability of a particle to be at $x$, irrespective of its status (alive or dead), corresponds to the combined probabilities of two types of trajectories: those that persist until the final time $t$ and are located at $x$, and those that die at $x$ before the final time, 
\begin{equation}  P_t(x)=P^{(a)}_t(x)+P^{(d)}_t(x).\label{eq:sticky split prob}
\end{equation}
A straightforward solution of \eqref{eqn:sticky pos eqns} gives
\begin{equation}
    P_t(x)= e^{-\alpha t}g_t(x) + \alpha \int _0^t  d{t'} \, e^{-\alpha t'} g_{t'}(x)
    \label{eqn: stickydistribution}
\end{equation}
where $g_t(x)=(4\pi t)^{-\frac{1}{2}}\exp(-x^2/4t)$ is the probability for a free Brownian starting at the origin to reach $x$ at time $t$. The first term in the solution is due to particles that remain alive throughout the entire duration, where $e^{-\alpha t}$ is the corresponding surviving probability. The second term is the net contribution from particles that survive until any intermediate time $t'<t$ reaching location $x$ and die in the interval between $t'$ and $t'+dt'$. 

For large time $t=T$ with $x=qT$, the first term on the right hand side of \eqref{eqn: stickydistribution} has an asymptotic $e^{-T (\alpha+q^2/4)}$, while the second term decays as $e^{-T\phi(q)}$ with
\begin{equation}
    \phi(q)= \begin{dcases} |q| \sqrt{\alpha} &0\leq |q|<\sqrt{4\alpha} \\
    \alpha+\frac{q^2}{4} & |q|\geq \sqrt{4\alpha}. \end{dcases}\label{eq:phi sticky 1}
\end{equation}
The latter asymptotic is obtained by introducing a change of variable $t'=r T$ and subsequently evaluating the integral in \eqref{eqn: stickydistribution} using a saddle point approximation for large $T$. The non-analyticity in \eqref{eq:phi sticky 1} arises because, for $\vert q\vert >\sqrt{4\alpha}$, the saddle-point value of $r=r^\star\equiv \vert q \vert/\sqrt{4\alpha}$ falls outside the range of integration. As a result, the integration is dominated by the value of the integrand at $r=1$ where it is maximum.

\begin{figure}
    \centering\includegraphics[width=0.48 \textwidth]{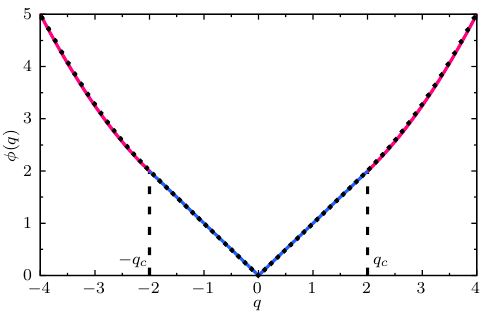}
    \caption{(color online) The solid line denotes the ldf in \eqref{eq:phi sticky 1} for the distribution of position of a mortal Brownian particle with a death rate $\alpha=1$. The change of color at $q=\pm q_c$ with $q_c=\sqrt{4\alpha}$ highlights the second-derivative-discontinuity of $\phi(q)$ in \eqref{eq:phi sticky 1}. The dots on the plot represent results obtained by importance sampling simulations for $T=100$ and $dt=0.01$ (see Appendix~\ref{app:importance sampling}). }
    \label{fig:one_sticky}
\end{figure}

For large $T$, the asymptotic of $P_T(x)$ is determined by the dominant contribution among the two terms in \eqref{eqn: stickydistribution} leading to \eqref{eqn:largedeviation} for $Q\equiv x$
with the ldf $\phi(q)$ in \eqref{eq:phi sticky 1}. Note that the position $x_t$ is an empirical observable \eqref{eqn:residencetime} with $U=0$ and $V=1$. The corresponding ldf is plotted in Fig.~\ref{fig:one_sticky} revealing singularities at $q=\pm \sqrt{4\alpha}$, where $\phi''(q)$ is discontinuous. Additionally, there is an extra singularity at $q=0$, where the first derivative $\phi'(q)$ has a jump discontinuity. Drawing an analogy with equilibrium phase transitions, we categorise the former type of singularities as second-order DPTs and the latter type as first-order DPTs. 

The origin of the second-order transition becomes apparent through the saddle point approximation used in deriving \eqref{eq:phi sticky 1}. In this approximation, discussed immediately after \eqref{eq:phi sticky 1}, the re-scaled variable $r$ denotes the fraction of time the particle remains alive. For $\vert q \vert<\sqrt{4\alpha}$, the dominant contribution comes from the trajectories that reach $x$ and die at time $T\sqrt{4\alpha}$. In contrast, for $\vert q\vert \ge \sqrt{4\alpha}$, the dominant contribution arises from trajectories that remain alive until the final time $T$. This drastic shift in the nature of most probable trajectories gives rise to the second-order dynamical phase transition.

The first-order transition at $q=0$ stems from a similar change in the nature of most probable trajectories.  For small $q$, the most probable trajectory is a straight line from the origin to $x=q T$, reaching in time $T r^\star=T \vert q \vert /\sqrt{4\alpha}$ and then dying. The slope of the straight line $x/(T r^\star)=\textrm{sgn}(q)\sqrt{4\alpha}$ is non-vanishing for $q\to 0$, but it changes discontinuously as $q$ changes sign, leading to a first-order transition.   

The ldf expression in \eqref{eq:phi sticky 1} reveals that for $\vert q \vert <\sqrt{4\alpha}$, the asymptotic probability $P_T(x=qT)$ in \eqref{eqn:largedeviation} is time independent. This implies, for distances $\vert x \vert <T\sqrt{4\alpha}$, the distribution of position reaches a stationary state.

The same ldf in \eqref{eq:phi sticky 1} was reported in the distribution of position $P_T(x)$ of a resetting Brownian motion \cite{majumdar2015dynamical}. This is not surprising, considering that the trajectories for $P_T(x)$ in the resetting Brownian motion after the last reset are related to the corresponding living part of the trajectories for the mortal Brownian motion by a reversal of time. This relation extends to other variants of the two dynamics, such as their respective generalisations for non-Markovian processes like fractional Brownian motion (see \sref{sec:fBm}).

The solution in \eqref{eqn: stickydistribution} follows
\begin{equation}
    \partial_t P_t(x)=\partial_{x}^2 P_t(x) - \alpha P_t(x)+\alpha \delta(x)\label{eq:full prob eq mortal BM}
\end{equation}
with the initial condition $P_0(x)=\delta(x)$. The same equation was observed for the resetting Brownian motion \cite{majumdar2015dynamical}, where the interpretation of the terms in \eqref{eq:full prob eq mortal BM} is clear.
For the mortal Brownian motion studied here, \eqref{eq:full prob eq mortal BM} can be interpreted
in a slightly different way. We note that $P_t(x)$ is the probability density of the mortal walker to be at $x$ at time $t$. Now, we split the time interval $[0,t+dt]$
into two segments $[0,dt]$ and $[dt,t+dt]$. In the initial interval $dt$, the particle, starting at
the origin, jumps to $ dx$ with probability $(1- \alpha dt)$ and dies (and therefore stays at $x=0$)
with the complementary probability $\alpha dt$. For the second interval $[dt, t+dt]$, the particle has to reach $x$ starting at $d x$ in time $t$, which is equal
to the probability that it reaches $x-d x$ in time $t$ starting at the origin. Considering these contributions, one can write the backward evolution equation of $P_t(x)$ as
\begin{equation}
P_{t+dt}(x)= (1- \alpha dt) \langle P_t(x-d x) \rangle + \alpha dt \delta(x)
\end{equation}
where $\langle \rangle$ denotes an average over the initial jump $d x= \eta(0) dt$, where $\eta(0)$ is the initial noise. Taylor expanding $P_t(x-dx)$ and using the fact that $\langle \eta(0)\rangle =0$
and $\langle \eta^2(0)\rangle = \frac{2}{dt}$, we recover \eqref{eq:full prob eq mortal BM} in the limit $dt\to 0$.

\subsection{Area under the trajectory \label{sec:area}}
\begin{figure}
    \centering\includegraphics[width=0.48 \textwidth]{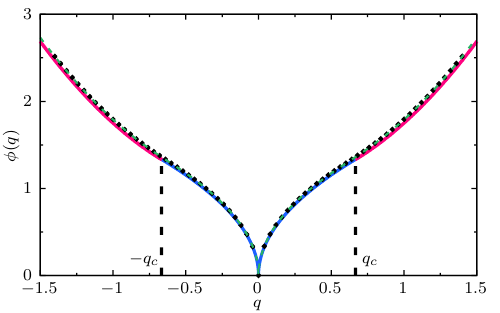}
    \caption{(color online) The ldf in \eqref{eq:phi sticky area} for the area under alive trajectories of a mortal Brownian particle with a death rate $\alpha=1$. The change of color at $q=\pm q_c$, where $q_c=\frac{2}{3}\sqrt{\alpha}$, signifies the second-derivative-discontinuity of $\phi(q)$ in \eqref{eq:phi sticky 1}. The data points indicate results from importance sampling simulations for $T = 100$ with $dt = 0.01$ (see Appendix.~\ref{app:importance sampling}). They slightly deviate from the result of \eqref{eq:phi sticky area} because the asymptotic limit $T\to\infty$ has not been reached yet, see the dashed green line corresponding to the exact ldf computed from \eqref{eqn: stickydistribution_area}.}
    \label{fig:one_sticky area}
\end{figure}

Another interesting observable is the area covered by a trajectory \cite{majumdar2005brownian}, defined as $Q=\int_0^{t_f}X_t dt$, where $t_f$ is the time at which the particle died. Similar to \eqref{eq:sticky split prob}, the probability $P_t(Q)$ measured at time $t$ can be written as a combination of two parts
\begin{equation}
P_t(Q)=P^{(a)}_t(Q)+P^{(d)}_t(Q)\label{eq:sticky split prob area}
\end{equation}
where $P^{(a)}_t(Q)$ is the contribution from trajectories that are alive until the final time $t$, whereas the trajectories contributing to $P^{(d)}_t(Q)$ have died earlier. It is straightforward to show that
\begin{equation}
    P_t(Q)= e^{-\alpha t}\widehat{g}_t(Q) + \alpha \int _0^t  d{t'} \, e^{-\alpha t'} \widehat{g}_{t'}(Q)
    \label{eqn: stickydistribution_area}
\end{equation}
where 
\begin{equation}
\widehat{g}_t(Q)=\left(\frac{4}{3}\pi t^3\right)^{-1/2}e^{-\frac{3Q^2}{4t^3}}
\end{equation}
is the probability of area $Q$ for a free Brownian particle (simple to derive using the linearity of the observable).
Following a similar saddle point analysis as used for the position distribution, we find that for large time $t=T$, $P(Q=qT^{2})\sim e^{-T\phi(q)}$, where the ldf is given by:
\begin{equation}
    \phi(q)= \begin{dcases} \sqrt{|q|} \, \frac{2\sqrt{2}}{\sqrt{3}}\alpha^{3/4} & \textrm{for } 0\leq |q|<\frac{2}{3}\sqrt{\alpha} \\
    \alpha+\frac{3}{4}q^2 & \textrm{for } |q|\geq \frac{2}{3}\sqrt{\alpha}. \end{dcases}\label{eq:phi sticky area}
\end{equation}
This scaling behaviour differs from the distribution in \eqref{eqn:largedeviation}. For similar examples, refer to \cite{Nickelsen2018,krapivsky2014melting}. The singular ldf is depicted in \fref{fig:one_sticky area}, which, for small $q$, grows as $\sqrt{q}$ compared to the linear growth in \eqref{eq:phi sticky 1}.

\subsection{A general class of observables}\label{sec:sticky general laplace}
The origin of the singularities in the ldfs for position and area lies in the competition between trajectories that remained alive for the entire duration and trajectories that died at an intermediate time. Such competition could result in similar singularities for other path-dependant observables. To illustrate this aspect, we consider a subset of observables $Q$ in \eqref{eqn:residencetime} with $V(X_t)=0$ while 
$U(X_t)$ is a function of the alive part of trajectory $X_t$:
\begin{equation}
     Q=\int_0^{t_f}dt\; U(X_t)\label{eq:sticky A}
\end{equation}
where $t_f$ is the time until which the particle is alive. { The choice of observables with  $V(X_t)=0$ is made solely for simplicity; the method applies equally well to nonzero $V(X_t)$.}

We define $P_T(Q\vert x_0)$ as the probability of the observable $Q$ at time $T$ for a mortal Brownian motion starting at $x_0$. Analogous to \eqref{eq:sticky split prob}, the contributions to $P_T(Q\vert x_0)$ are split into two classes of trajectories:
\begin{equation}
    P_T(Q\vert x_0)=P^{\text{(a)}}_T(Q\vert x_0)+P^{\text{(d)}}_T(Q\vert x_0)
    \label{eq:sticky BM prob first}
\end{equation}
where $P^{\text{(a)}}_T(Q\vert x_0)$ is the contribution from the living trajectories, while $P^{\text{(d)}}_T(Q\vert x_0)$ is the contribution from trajectories that died before time $T$. The two probabilities are related,
\begin{equation}
    P^{\text{(d)}}_T(Q\vert x_0)=\alpha \int_0^Tdt P^{\text{(a)}}_t(Q\vert x_0).
    \label{eq:relation Pd & Pa}
\end{equation}
For determining $P^{\text{(a)}}_t(Q\vert x_0)$, we adopt a backward Fokker-Planck equation approach discussed in \cite{majumdar2005brownian}. We define a double Laplace transformation 
\begin{equation}
    \mathcal{D}_{t\to s}^{Q\to p}\odot f_t(Q) =\int_0^\infty dt e^{-st}\int dQ \; e^{p Q} f_t(Q).
    \label{eq:double L transform}
\end{equation}
Expressing \eqref{eq:sticky BM prob first} in terms of the transformed variables and using \eqref{eq:relation Pd & Pa} gives
\begin{equation}
    R_s(p, x_0)=S_s(p, x_0)+\frac{\alpha}{s}S_s(p,x_0)
    \label{eq:sticky BM Laplace first}
\end{equation}
where $R_s(p,x_0)$ and $S_s(p,x_0)$ are the $\mathcal{D}$-transform of $P_T(Q\vert x_0)$ and $P^{\text{(a)}}_T(Q\vert x_0)$.

In line with the methods illustrated in \cite{majumdar2005brownian}, we find that $S_s(p,x_0)$ is a solution of the ordinary differential equation \begin{subequations}
\begin{equation}
    \frac{d^2 S_s}{dx_0^2}-(\alpha+s-pU(x_0))S_s=-1.
\end{equation}
The specific boundary condition hinges on the observable in question. For observables where $U(x)\to 0$ for large $x$,
\begin{equation}
    S_s(p,x_0)=\frac{1}{\alpha+s} \qquad \textrm{for }x_0\to \pm \infty.
\end{equation}\label{eq:sticky eq for S}\end{subequations}

In principle, obtaining a solution for $S_s(p,x_0)$ allows us to compute $R_s(p,x_0)$ in \eqref{eq:sticky BM Laplace first} and perform the inverse $\mathcal{D}$-transformation to derive the probability $P_T(Q\vert x_0)$. However, performing these steps analytically for arbitrary $U(x)$ is not always feasible. Nevertheless, a large deviation form of the probability $P_T(Q\vert x_0)$ can be deduced from the poles of $R_s(p,x_0)$. This can be seen as follows. Assuming $P_T(Q=qT\vert x_0)\sim e^{-T\phi(q)}$, the Laplace transformation
\begin{equation}
    \int dQ \, e^{pQ}\, P_T(Q=qT)\sim e^{T\mu(p)}
\end{equation}
for large $T$, where $\mu(p)$ is the scgf related to $\phi(q)$ by \eqref{eqn:legendre}. An additional Laplace transformation gives
\begin{equation}
    R_s(p,x_0)\sim \frac{1}{s-\mu(p)}.
\end{equation}
Therefore, a pole of $R_s(p,x_0)$ at $s=\mu(p)$ suggests a large deviation asymptotic of $P_T(Q=qT\vert x_0)$ where the ldf $\phi(q)$ is a Legendre-Fenchel transform \eqref{eqn:legendre} of $\mu(p)$. If $R_s(p,x_0)$ has multiple poles, the scgf $\mu(p)$ is given by the maximum among all the poles.

{ Similar pole structure is seen in the calculation of the ldf of $P_t(x)$ in \eqref{eq:phi sticky 1}, corresponding to $Q=x$ for which $U=0$ and $V=1$ in \eqref{eqn:residencetime}. Referring to \eqref{eqn: stickydistribution}, it is straightforward to demonstrate that the corresponding $R_s(p,0)=\frac{\alpha +s}{s \left(\alpha -p^2+s\right)}$, which has two poles at $s=0$ and at $s=p^2-\alpha$. (Here, $x_0=0$ is the initial position for this problem.) This results in the scgf, $\mu(p)=\max\{0,p^2-\alpha\}$; and its Legendre-Fenchel transform yields the ldf in \eqref{eq:phi sticky 1}.}

In the following, we employ this general approach to obtain large deviations function for a couple of observables with singular ldf. 

\textit{Remark:} A crucial observation in this problem is that the slope $\sqrt{\alpha}$ of the linear part of the ldf in \eqref{eq:phi sticky 1} for $\vert q\vert <2\sqrt{\alpha}$ corresponds to the pole of $s R_s(p,0)$ with respect to $p$ for vanishing $s$. This structure is more general. For an arbitrary $Q$, if the $P_T(Q\vert x_0)$ is stationary within a range of $Q$ such that the ldf $\phi(q)=c q$, then the corresponding $sR_s(p,x_0)$ for vanishing $s$ has a pole at $p=c$.

\textit{Remark:} {For the observable Area in \sref{sec:area}, the distribution follows an asymptotic $P(Q=q T^2)\sim e^{-T\phi(q)}$ and the generating function $\langle e^{p Q}\rangle \sim e^{T\mu(p T)}$, for large $T$. It is straightforward to incorporate such examples within our theory by scaling $p=\hat{p}/T$.}

\subsubsection{Local time}
For $U(x)=\delta(x-a)$, the observable $Q$ corresponds to the local time density \cite{pal2019local,burenev2023local}, where $Q dx$ gives the net amount of time spent by the alive particle in the window $[a-\frac{dx}{2},a+\frac{dx}{2}]$ up to time $T$.
\begin{figure}
    \centering
    \includegraphics[width=0.48 \textwidth]{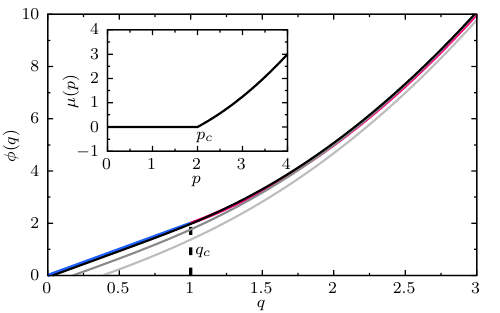}
    \caption{(color online) The blue–pink curve shows the ldf in \eqref{eq:ldf local mortal} for the local time at the origin of a mortal Brownian particle with death rate $\alpha=1$, starting at the origin. The linear part (colored blue) meets the non-linear part (colored pink) at a singular point $q_c=1$.  Uniformly colored curves correspond to ldfs extracted from the exact distribution $P_T(Q\vert 0)$ expressed in terms of the scaling function $h(x,t)$ in \eqref{eq:exact h}, for $T=1$ (lightest), $T=2$ (intermediate), and $T=10$ (darkest). The inset shows the associated scgf from from \eqref{eq:scgf local time mortal BM}, with singularity at $p_c=2$.}
    \label{fig:sticky_local_scgf}
\end{figure}

The solution for \eqref{eq:sticky eq for S} in this case is given by
\begin{equation}
    S_s(p,x_0)=\frac{1}{\alpha+s}\left\{ 1-\frac{p}{p-2\sqrt{\alpha+s}}e^{-\vert x_0-a\vert \sqrt{\alpha+s}}\right\}.
\end{equation}
Substituting this into \eqref{eq:sticky BM Laplace first} reveals that $R_s(p,x_0)$ has two poles, at $s=0$ and at $s=\frac{1}{4}p^2-\alpha$. Consequently
\begin{equation}
    \mu(p)=\max\left\{0,\frac{p^2}{4}-\alpha\right\}.\label{eq:scgf local time mortal BM}
\end{equation}
The scgf in \eqref{eq:scgf local time mortal BM} has a singularity at $p=2\sqrt{\alpha}$. Performing a Legendre-Fenchel transformation of the scgf results in a singular ldf given by
\begin{equation}
    \phi(q)= \begin{dcases} 2q \sqrt{\alpha} &0\leq q<\sqrt{\alpha}, \\
    \alpha+q^2 & q\geq \sqrt{\alpha}. \end{dcases}\label{eq:ldf local mortal}
\end{equation}
This singularity is explicitly verified for $x_0=a$, for which the inverse-$\mathcal{D}$ transformation of $R_s(p,x_0)$ can be analytically performed, yielding
$P_T(Q\vert a)=\frac{1}{\sqrt{T}}h\left(\frac{Q}{\sqrt{T}},\sqrt{\alpha T}\right)$, where
\begin{align}\label{eq:exact h}
    h(x,y)=&\frac{2e^{-x^2-y^2}}{\sqrt{\pi}}+\nonumber\\&y\left(e^{-2xy}\textrm{erfc}(x-y)-e^{2xy}\textrm{erfc}(x+y)\right).
\end{align}
{The ldf \eqref{eq:ldf local mortal} and the corresponding scgf \eqref{eq:scgf local time mortal BM} are shown in Fig.~\ref{fig:sticky_local_scgf}, with a comparison with the exact result \eqref{eq:exact h} for increasing values of $T$.}

\subsubsection{Residence time}\label{sec:stickyresidencelaplace}

Next example we consider is the time spent by the particle within an interval while it was alive. To keep the algebra simple, we consider a symmetric interval $-a\le x\le a$ with respect to the starting point $x_0=0$. The residence time is the observable \eqref{eqn:residencetime} with $U(x)=1$ inside this interval and zero outside \cite{burenev2023occupation}.

The solution to \eqref{eq:sticky eq for S} for this specific $U(x)$ is expressed as follows: for $x\in [-a,a]$,
\begin{equation}
    S_s(p,x_0)=\frac{1}{\widehat{p}}\left(-1+p \frac{\cos(x_0\sqrt{\widehat{p}})}{d(s,p)\sqrt{\widehat{s}}}\right)
\end{equation}
and outside this region
\begin{equation}
    S_s(p,x_0)=\frac{1}{\widehat{s}}\left(1+p \frac{\sin(a\sqrt{\widehat{p}})e^{(a-\vert x_0\vert)\sqrt{\widehat{s}}}}{d(s,p)\sqrt{\widehat{p}}}\right)
\end{equation}
where we denote $\widehat{p}=p-s-\alpha$,  $\widehat{s}=s+\alpha$, and the denominator
\begin{equation}
d(s,p)=\sqrt{\widehat{s}}\cos(a\sqrt{\widehat{p}})-\sqrt{\widehat{p}}\sin(a\sqrt{\widehat{p}}).\label{eq:dsp}
\end{equation}
Using this solution, the expression for $R_s(p,x_0)$ in \eqref{eq:sticky BM Laplace first} can be written as
\begin{equation}
    R_s(p,x_0)=\frac{n(s,p,x_0)}{s\;d(s,p)}\label{eq:R pole residence sticky}
\end{equation}
where the numerator is analytic in $s$, and the non-trivial poles for $R_s(p,x_0)$ arise from the roots of $d(s,p)$.

\begin{figure}
    \centering
    \includegraphics[width=0.48 \textwidth]{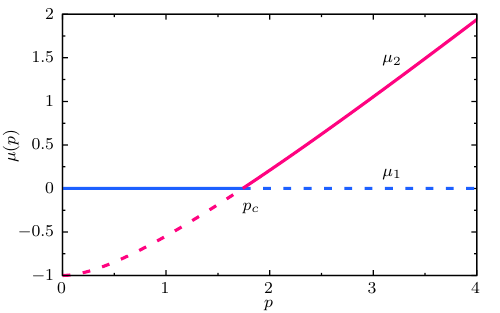}
    \caption{(color online) The solid line represents the scgf in \eqref{eq:cgf sticky residence} for the residence time in the interval $[-1,1]$ of a mortal Brownian particle with a death rate $\alpha=1$. The scgf is the maximum of the two poles: $\mu_1=0$ (blue) and $\mu_2(p)=s^\star(p)$ (pink) of $R_s$ in \eqref{eq:R pole residence sticky}, {leading to a singularity at $p_c=1.74$. The pink curve meets y-axis at value $\mu_2(0)=-\alpha$ which can be seen from the largest root $s^\star$ of $d(s,0)$ in \eqref{eq:dsp}.}}
    \label{fig:sticky_residence_scgf}
\end{figure}
\begin{figure}
    \centering
    \includegraphics[width=0.48 \textwidth]{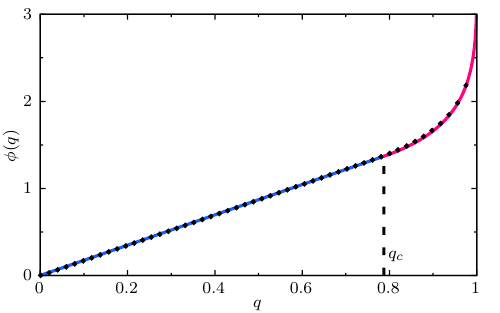}
    \caption{(color online) The ldf for residence time of a mortal Brownian particle with the same parameter values of \fref{fig:sticky_residence_scgf}. The solid line represents the theoretical result in \eqref{eq:ldf sticky residence} while the points indicate results from importance sampling simulations with parameter values $T = 100$, $dt = 0.01$ (see Appendix.~\ref{app:importance sampling}). At the singular point $q_c=0.78$, marked by the dashed line, the $\phi''(q)$ is discontinuous.}
    \label{fig:sticky_residence_ratefn}
\end{figure}

Following our argument discussed earlier, the scgf $\mu(p)$ is the largest non-negative root in $s$ for which $s d(s,p)=0$. It turns out that for $p$ less than a critical value $p_c$, defined as the positive solution of $d(0,p_c)=0$, there are no positive roots $s^\star$ such that $d(s^\star,p)=0$. In this scenario, the trivial pole $s=0$ in \eqref{eq:R pole residence sticky} gives the scgf $\mu(p)=0$. Altogether, this results in a singular scgf
\begin{equation}
    \mu(p)= \begin{dcases} 0 & \textrm{for } 0\leq p<p_c \\
    s^\star(p) & \textrm{for } p_c\geq p. \end{dcases}\label{eq:cgf sticky residence}
\end{equation}
with $s^\star(p)$ and $p_c$ defined above.

The Legendre-Fenchel transformation \eqref{eqn:legendre} of the scgf in \eqref{eq:cgf sticky residence} results in the singular ldf
\begin{equation}
    \phi(q)= \begin{dcases} p_c q  &\textrm{for }0\leq q<q_c, \\
    \max\{pq-s^\star(p)\} &\textrm{for } q\geq q_c, \end{dcases}\label{eq:ldf sticky residence}
\end{equation}
where $q_c=\partial_p s^\star(p_c)$. The scgf \eqref{eq:cgf sticky residence} and the ldf \eqref{eq:ldf sticky residence} are plotted in \fref{fig:sticky_residence_scgf} and \fref{fig:sticky_residence_ratefn}, respectively, for specific parameter values.

\section{Brownian Motion in presence of an absorbing boundary \label{sec:BM abs}}
In this second example, we examine a one-dimensional Brownian particle confined within an absorbing wall at the origin and a reflecting wall at $x=L$. As illustrative examples, we focus on a sub-class of path dependent observables given by \eqref{eqn:residencetime}, where $V(x)=0$ for all $x$.

Similar to \eqref{eq:sticky BM prob first}, the contribution to the probability $P_T(Q\vert x_0)$ of $Q$ for the particle starting at $x_0$ arises from two classes of trajectories:
\begin{equation}
    P_T(Q\vert x_0)=P^{\text{sur}}_T(Q\vert x_0)+\int_0^Tdt P^{\text{fp}}_t(Q\vert x_0),
    \label{eq:abs BM prob first}
\end{equation}
where $P^{\text{sur}}_T(Q\vert x_0)$ is the joint probability for the particle to survive up to time $T$ without being absorbed and contribute $Q$, while $P^{\text{fp}}_t(Q\vert x_0)$ is the probability for the particle to get absorbed at time $t<T$ and contribute an amount $Q$ to the observable.

Applying double Laplace transformation \eqref{eq:double L transform} to
\eqref{eq:abs BM prob first} yields
\begin{equation}
    R_s(p, x_0)=S_s(p, x_0)+\frac{1}{s}F_s(p,x_0),
    \label{eq:abs BM Laplace first}
\end{equation}
where the terms are the $\mathcal{D}$-transform of respective probabilities in \eqref{eq:abs BM prob first}.
Utilising a backward Fokker-Planck approach described in \cite{majumdar2005brownian}, we find that $S_s(p,x_0)$ is a solution of the ordinary differential equation \begin{subequations}
\begin{equation}
    \frac{d^2 S_s}{dx_0^2}-(s-pU(x_0))S_s=-1
\end{equation}
with the boundary conditions 
\begin{equation}
    S_s(p,0)=0\quad \textrm{and }\frac{d S}{dx_0}\bigg\vert_{x_0=L}=0.
\end{equation}\label{eq:eq for S}\end{subequations}\begin{subequations}
Similarly, $F_s(p,x_0)$ satisfies the following ordinary differential equation:
\begin{equation}
    \frac{d^2 F_s}{dx_0^2}-(s-pU(x_0))F_s=0
\end{equation}
with the boundary conditions
\begin{equation}
    F_s(p,0)=1\quad \textrm{and }\frac{d F_s}{dx_0}\bigg\vert_{x_0=L}=0.
\end{equation}\label{eq:eq for F}\end{subequations}
As we discussed in \sref{sec:sticky general laplace}, the presence of a pole for $R_s(p,x_0)$ at $s=\mu(p)$ implies a large deviation form of the probability $P_T(Q\vert x_0)$, where the ldf $\phi(q)$ relates to the scgf $\mu(p)$ through a Legendre-Fenchel transformation \eqref{eqn:legendre}. 

In the following, we apply this general approach to derive ldfs for two observables considered in \sref{sec:sticky general laplace}. 

\subsection{Local time}\label{sec:localabsorbing}
For $U(x)=\delta(x-a)$ with $0<a<L$, the observable $Q$ in \eqref{eqn:residencetime} is the local time density.  
For this observable, the solutions for $S_s$ and $F_s$ in (\ref{eq:eq for S},\,\ref{eq:eq for F}) are straightforward and are expressed as
\begin{equation}
    S_{u^2}(p,x_0)=\frac{n_1(u,p,x_0)}{d(u,p)},\quad     F_{u^2}(p,x_0)=\frac{n_2(u,p,x_0)}{d(u,p)} \label{eq:sticky local S and F}
\end{equation}
where both the numerators (expressions given in the appendix \ref{app:expressions}) and the common denominator
\begin{equation}
    d(u,p)=p \sinh \left(a u\right) \cosh \left(u (L-a)\right)+u \cosh
   \left(L u\right)\label{eq:abs denominator d}
\end{equation}
are analytic functions of $u$. Using these solutions in \eqref{eq:abs BM Laplace first}, we obtain
\begin{equation}
    R_{u^2}(p,x_0)=\frac{u^2 n_1(u,p,x_0)+n_2(u,p,x_0)}{u^2 d(u,p)}.\label{eq:R for local abs}
\end{equation}

For $p$ exceeding a critical value $p_c=1/a$, the denominator $d(u,p)$ has real roots at non-zero $u=\pm u_c(p)$. Consequently, $R_{s}(p,x_0)$ features a pole at positive $s=u_c(p)^2$ for $p>p_c$.

\begin{figure}
\includegraphics[width=0.48 \textwidth]{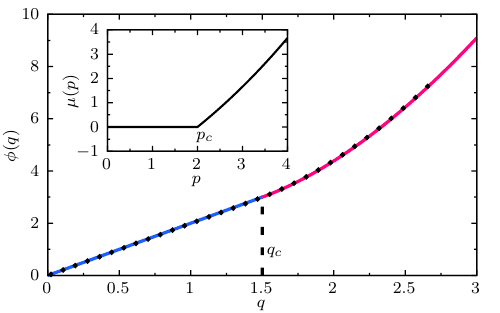}%
\caption{(color online) The ldf \eqref{eq:scgf ldf for local abs} for local time at $a=1/2$ for a Brownian particle confined between an absorbing wall at the origin and a reflecting wall at $L=1$. At the singular point $q_c=1.5$, marked by the dotted line, the $\phi''(q)$ is discontinuous. The data points indicate results from importance sampling simulations for $T = 100$ with $dt = 0.001$ (see Appendix.~\ref{app:importance sampling}). The inset shows the corresponding scgf given in \eqref{eq:scgf residence abs} which is singular at $p_c=2$.}
\label{fig:wall_local_time_ratefn}
\end{figure}

For $p<p_c$, the denominator $d(u,p)$ has no real roots in $u$, and consequently, $R_s(p,x_0)$ lacks a pole at $s>0$. In this case, the leading contribution to the scgf arises from the real pole at $s=0$. 

Altogether, this yields the piece-wise scgf 
\begin{subequations}\begin{equation}
    \mu(p)=\begin{dcases}
    0 & p <1/a \\
    u_c(p)^2 & p\geq 1/a
    \end{dcases}\label{eq:scgf residence abs}
\end{equation}
where $u_c$ is the largest real solution of the transcendental equation $d(u_c,p)=0$ in \eqref{eq:abs denominator d}.

A Legendre-Fenchel transformation \eqref{eqn:legendre} leads to the ldf
\begin{equation}
    \phi(q)=\begin{dcases}
    \frac{1}{a}q & q <q_c \\
    \max_p\{qp -u_c^2(p)\} & q\ge q_c
    \end{dcases}\label{eq:scgf ldf for local abs}
\end{equation}\end{subequations}
with $q_c=\mu^\prime(p_c^+)$ where the $^+$ indicates that the derivative is evaluated on the right side of $p_c$. The singular ldf and the corresponding scgf are plotted in \fref{fig:wall_local_time_ratefn}. 

{\textit{Remark:}} The linear part of ldf for $q<q_c$ implies a stationary regime for the $P_T(Q\vert x_0)$. As discussed in \sref{sec:sticky general laplace}, the slope $1/a$ for the ldf in this regime can also be observed from the pole of $s R_s(p,x_0)$ at $p=1/a$ for vanishing $s$.

\subsection{Residence time}\label{sec:BM abs res laplace}
\begin{figure}
\includegraphics[width=0.48 \textwidth]{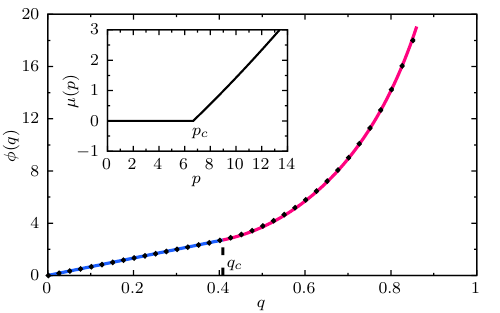}%
\caption{(color online) The ldf for residence time of a Brownian particle in the interval $[1/3,2/3]$, confined between an absorbing wall at the origin and a reflecting wall at $L=1$. At the singular point $q_c=0.408$, marked by the dotted line, the $\phi''(q)$ is discontinuous. The data points indicate results from importance sampling simulations for $T = 20$ with $dt = 0.001$ (see Appendix.~\ref{app:importance sampling}). The inset shows the corresponding scgf with singularity at $p_c=6.72$.}
\label{fig:wall_time_ratefn}
\end{figure}

The residence time is defined by the observable $Q$ in \eqref{eqn:residencetime} with $V(x)=0$ for all $x$, while $U(x)=1$ in the interval $x\in[a,b]$ for $0<a<b<L$, zero outside this interval. 
The solution for the corresponding $S$ and $F$ in (\ref{eq:eq for S},\,\ref{eq:eq for F}) is straightforward but cumbersome to write in the main text. (For their explicit solution, refer to the supplemental Mathematica notebook \cite{extra}.) The solution follows the structure of \eqref{eq:sticky local S and F}, with a denominator
\begin{equation}
\begin{split}
    &d_{\text{res}}(u,p)=\left(p-2u^2\right) \tanh \left(u (a-b+L)\right)\\
    &+\frac{p\sinh \left(u (a+b-L)\right)}{\cosh \left(u (a-b+L)\right)}+\frac{2 u\sqrt{ (p-u^2)}}{ \tan ((a-b) \sqrt{p-u^2})}.
    \end{split}
    \label{eq:abs denominator d residence time}
\end{equation}
The subsequent analysis parallels that of the local time discussed in the previous section. The scgf has similar formal structure as in \eqref{eq:cgf sticky residence}, where $u_c$ is the largest positive root of $d_{\text{res}}(u,p)$. This root exists above a non-zero threshold value of $p_c$, which is obtained by $d_{\text{res}}(0,p_c)=0$ (refer to the discussion above \eqref{eq:cgf sticky residence}). The resulting scgf and the corresponding ldf are shown in  \fref{fig:wall_time_ratefn}.

\textit{Remark:} Singular scgf is not limited to the finite domain of the Brownian motion. When the reflecting boundary is pushed to infinity ($L\to \infty$), from \eqref{eq:abs denominator d residence time} we get that $\mu(p)$ is the largest non-negative solution $s$ of the transcendental equation
\begin{equation}
     \frac{2\sqrt{s( p-s)}}{\tan \left((a-b) \sqrt{p-s}\right)}=2s-p\left(1-e^{-2a\sqrt{s}}\right)
\end{equation}
Resulting scgf and the corresponding ldf have similar singular structure as in \fref{fig:wall_time_ratefn}. Effect of conditioning on Brownian motion in a similar geometry has been recently studied in \cite{Monthus_2022}.

\section{Spectral method}
\label{sec:tiltedintro}
The problem of the long-time asymptotics \eqref{eqn:ratefunction} of the probability $P(Q)$ of empirical observables \eqref{eqn:residencetime} can be approached using an alternative method that has recently gained considerable attention \cite{jack2010large, jack2015effective,chetrite2013nonequilibrium,chetrite2015nonequilibrium,derrida2019large}. In this approach, large deviations are expressed in terms of largest eigenvalue of an operator, offering an intuitive analogy with free-energy calculations using the transfer matrix method \cite{Goldenfeld,Dhar2011,cuesta2004general}.

We present a brief introduction to this method using the simplest example of a standard Brownian motion $\dot{X}_t=\eta_t$ with Gaussian white noise $\eta_t$ of mean zero and covariance of  $\langle\eta_t\eta_{t'}\rangle=2\delta(t-t')$. For the joint probability $P_T(Q,x_T\vert x_0)$ of the observable $Q$ in \eqref{eqn:residencetime} and the final position $x_T$, starting at $x_0$, the generating function 
\begin{equation}
    R_T(p, x_T\vert x_0)=\int dQ e^{pQ}P_T(Q,x_T\vert  x_0)\label{eq:R P relation at T}
\end{equation}
has a path integral representation \cite{majumdar2005brownian},
\begin{equation}
    R_T(p,x_T\vert x_0)=\int_{x_0}^{x_T} \mathcal{D}[x_t]e^{-S_T(x_t)}
\end{equation}
where the Feynman-Kac action
\begin{equation}
    S_T(x_t)=\int_0^Tdt\,\left\{\frac{\dot{x}_t^2}{4}-p \dot{x}_tV(x_t)-p U(x_t)\right\}
\end{equation}
is analogous to the action of a charged particle in an electromagnetic field. The corresponding analogue of the Schrodinger equation is the Feynman-Kac formula \cite{majumdar2005brownian}
\begin{subequations}
\begin{equation}
     \frac{d}{dT}R_T(p,x\vert x_0)=\mathcal{L}_p\cdot R_T(p,x\vert x_0)
\end{equation}
where the operator\label{eq:equation for R forward}
\begin{equation}
   \mathcal{L}_p=\left(\frac{d}{dx}-p V\right)^2 +pU.
\label{eqn:tilted}
\end{equation}\end{subequations}
is often referred to as the tilted operator \cite{jack2010large, jack2015effective,chetrite2013nonequilibrium,chetrite2015nonequilibrium,derrida2019large}.

If the operator $\mathcal{L}_p$ has a spectral gap between its largest and the second-largest eigenvalues, then for large $T$, $R_T(p,x_T\vert x_0)\simeq e^{T\mu(p)}r_p(x_T)\ell_p(x_0)$, where $\mu(p)$ is the largest eigenvalue of $\mathcal{L}_p$ with the associated left eigenvector $\ell_p(x)$ and right eigenvector $r_p(x)$ satisfying
\begin{equation}
    \mathcal{L}^\dagger_p \ell_p(x)=\mu(p)\ell_p(x) \textrm{ ~~ and } \mathcal{L}_p r_p(x)=\mu(p)r_p(x),\label{eq:eigenequation ell}
\end{equation}
respectively, where $\dagger$ denotes the adjoint. Comparing the asymptotics of $R_T(p,x_T\vert x_0)$ with the asymptotics \eqref{eqn:scgf} one can recognise that the largest eigenvalue $\mu(p)$ gives the scgf of $Q$.

For a strictly convex $\mu(p)$, the asymptotic for $R_T(p,x_T\vert x_0)$ corresponds \cite{derrida2019large} to the large $T$ asymptotic of the joint probability \begin{subequations}\label{eq:PTxT together}
\begin{equation}
    P_T(Q=qT,x_T\vert x_0)\simeq \sqrt{\frac{\phi''(q)}{2\pi T }}e^{-T\phi(q)}r_p(x_T)\ell_p(x_0)\label{eq:asymptotic of PT}
\end{equation}
with
\begin{equation}
    \phi(q)=p q -\mu(p)\quad \textrm{and } \mu'(p)=q, \label{eq:legendre explicit}
\end{equation}\end{subequations}
which is seen by a saddle point approximation of the integral in \eqref{eq:R P relation at T}. 

The relation \eqref{eq:legendre explicit} is an explicit version of \eqref{eqn:legendre} for a strictly convex $\mu(p)$. It establishes an equivalence between the conditioned ensemble and the ensemble where evolution is weighted by the value of the observable. Such equivalence of ensemble also extends to the dynamics \cite{jack2010large, jack2015effective,chetrite2013nonequilibrium,chetrite2015nonequilibrium,derrida2019large}, more precisely, to the joint probability $P_t(Q,x\vert x_T,x_0)$ of being at location $x$ at an intermediate time $0<t<T$, starting at $x_0$ and ending at $x_T$, and yielding a value $Q$ for the empirical observable \eqref{eqn:residencetime} measured over the time window $[0,T]$. Analogous to \eqref{eq:PTxT together}, the joint probability can be estimated in the large time $T$ limit by the asymptotics of the corresponding generating function $G_t(p,x\vert x_T,x_0)$, defined in a way similar to \eqref{eq:R P relation at T}. This further implies \cite{jack2010large, jack2015effective,chetrite2013nonequilibrium,chetrite2015nonequilibrium,derrida2019large} that the conditional probability
\begin{equation}
    P_t(x\vert Q,x_T, x_0)=\frac{P_t(Q,x\vert x_T, x_0)}{\int dx P_t(Q,x\vert x_T, x_0)}
\end{equation}
can be estimated in the large $T$ limit by an analogous quantity in the weighted ensemble,
\begin{equation}
    P_t(x\vert Q=qT,x_T, x_0)\simeq \frac{G_t(p,x\vert x_T, x_0)}{\int dx G_t(p,x\vert x_T, x_0)}\label{eq:Pt Gt equivalence}
\end{equation}
with $p$ and $q$ related by \eqref{eq:legendre explicit}.

To describe the dynamics followed by the conditional probability $P_t(x\vert Q,x_T,x_0)$, we follow a procedure similar to that described in \cite{majumdar2015effective} for conditioned dynamics of a Brownian bridge. Using Markovianity, 
\begin{equation}
    G_t(p,x\vert x_T, x_0)= R_{T-t}(p,x_T\vert x)R_{t}(p,x\vert x_0)\label{eq:Gt markovian}
\end{equation}
where $R_t(p,x \vert x_0)$ evolves according to \eqref{eq:equation for R forward} with $T$ replaced by $t$, and $R_{T-t}(p,x_T\vert x)\equiv B_t(p,x\vert x_T)$ follows the backward evolution \cite{majumdar2005brownian}
\begin{equation}
    -\frac{d}{dt}B_t(p,x\vert x_T)=\mathcal{L}^\dagger \cdot B_t(p,x\vert x_T).
\end{equation}
It then requires a straightforward algebra to show that 
$G_t(p,x\vert x_T, x_0)$  in \eqref{eq:Gt markovian} satisfies
\begin{equation}
\frac{dG_t}{dt}=-2\frac{d}{dx}\left\{\left(p V(x) +\frac{1}{B_t}\frac{dB_t}{dx}\right)G_t\right\}+\frac{d^2G_t}{dx^2}.\label{eq:Gt evolution}
\end{equation}
Integrating the above equation over $x$, we see that the denominator in \eqref{eq:Pt Gt equivalence} is independent of time, assuming that $G_t(p,x\vert x_T,x_0)$ vanishes sufficiently quickly for $x\to \pm \infty$. Using these results in \eqref{eq:Pt Gt equivalence}, it immediately follows that the effective dynamics of the conditional probability $P_t(x\vert qT, x_T, x_0)$ is similar to \eqref{eq:Gt evolution}, with $p$ and $q$ related by \eqref{eq:legendre explicit}.

The effective dynamics particularly simplifies for $1\ll t \ll T$, where, for typical observables, the system reaches a quasi-stationary state \cite{jack2010large, jack2015effective,chetrite2013nonequilibrium,chetrite2015nonequilibrium,derrida2019large}, in which the conditional probability $P_t(x\vert qT, x_T,x_0)\simeq P_t(x\vert q)$ is independent of $x_T$, $x_0$, and $T$. Moreover, $B_t(p,x\vert x_T)\simeq e^{-t\mu(p)}\ell_p(x)r_p(x_T)$. The resulting conditioned dynamics is a drifted Brownian motion \cite{jack2010large, jack2015effective,chetrite2013nonequilibrium,chetrite2015nonequilibrium,derrida2019large} governed by the Fokker-Planck equation
\begin{subequations}\begin{equation}
\frac{d}{dt}P_t(x\vert q)=-\frac{d}{dx}\Big(F_t(x,p)P_t(x\vert q)\Big)+\frac{d^2}{dx^2}P_t(x\vert q)
\end{equation}
with a modified force
\begin{equation}
    F_t(x,p)=2\left\{p V(x) +\frac{\ell_p '(x)}{\ell_p(x)}\right\}
\end{equation}\label{eq:effective continume Langevin}\end{subequations}
where the left eigenvector $\ell_p(x)$ is defined in \eqref{eq:eigenequation ell}, and the parameters $p$ and $q$ are related by \eqref{eq:legendre explicit}. This effective description for the conditioned dynamics suitably generalizes \cite{jack2010large, jack2015effective,chetrite2013nonequilibrium,chetrite2015nonequilibrium,derrida2019large} for general diffusive processes and for Markov processes on discrete states and time.

In this spectral method, singularities in $\mu(p)$ can arise from crossing of the two leading eigenvalues, denoted as $\mu_1(p)$ and $\mu_2(p)$, of the tilted operator $\mathcal{L}_p$ in \eqref{eqn:tilted}, such that the scgf $\mu(p)=\max\{\mu_1(p),\mu_2(p)\}$. This is analogous to the origin of singular free energy at the phase transition point of equilibrium lattice models from the crossing of leading eigenvalues of the associated transfer matrix \cite{Goldenfeld,Dhar2011,cuesta2004general}. Across the singular point, the largest eigenvector also changes abruptly indicating a sudden change in the effective dynamics  \eqref{eq:effective continume Langevin}. This way the singularity of scgf $\mu(p)$ and equivalently that of the ldf $\phi(q)$ corresponds to a phase transition in the system's dynamics as the condition on the observable value is varied.

We illustrate how the DPTs for the models we discussed in \sref{sec:stickybrownian} and \sref{sec:BM abs} can be described using the spectral method. Our discussion will focus on the residence time for the two models; however, the analysis for the other observables follows a similar analysis.

\subsection{Residence time for a Brownian motion in presence of an absorbing wall}
\label{sec:residencecontinuous}
We initiate our discussion by revisiting the residence time problem, studied in \sref{sec:BM abs res laplace}, of a Brownian particle in presence of an absorbing and a reflecting walls at $x=0$ and $x=L$ respectively. Here, we reproduce our earlier results using the spectral method, with the observable under consideration being the residence time in the interval $x\in[a,b]$ with $0<a<b<L$.

From the previous section \eqref{eqn:tilted}, we see that the titled operator in this example takes the form 
\begin{equation}
    \mathcal{L}_p=\frac{d^2}{dx^2} +p\,\mathbb{1}_{[a,b]}.
\label{eqn:tiltedresidence}
\end{equation}
The scgf, $\mu(p)$, for the residence time is given by the largest eigenvalue of $\mathcal{L}_p$. Here, the operator has an eigenvalue $\mu_1(p)=0$ for all $p$ and an associated left eigenfunction denoted by $\ell_p^{(1)}(x)$. For $p=0$, this eigenvalue corresponds to the empty stationary state due to the absorbing wall, for which the left eigenfunction $\ell_0^{(1)}(x)=1$. Note that this eigenfunction does not vanish at the absorbing wall. The second relevant eigenvalue $\mu_2(p)$ is the largest eigenvalue of $\mathcal{L}_p$ with a vanishing boundary condition $\ell_p^{(2)}(0)=0$ for the corresponding eigenfunction $\ell_p^{(2)}(x)$ at the absorbing wall.
On the reflecting wall at $x=L$, both eigenfunctions, $\ell_p^{(1)}(x)$ and $\ell_p^{(2)}(x)$, satisfy the zero gradient condition $\ell_p^{(1)\prime}(L)=0 $ and $\ell_p^{(2)\prime}(L)=0$ \cite{du2020dynamical}. In Appendix \ref{sec:BCabsorbingappendix}, we discuss an origin of the absorbing boundary condition.
\begin{figure}
    \centering
    \includegraphics[width=0.48 \textwidth]{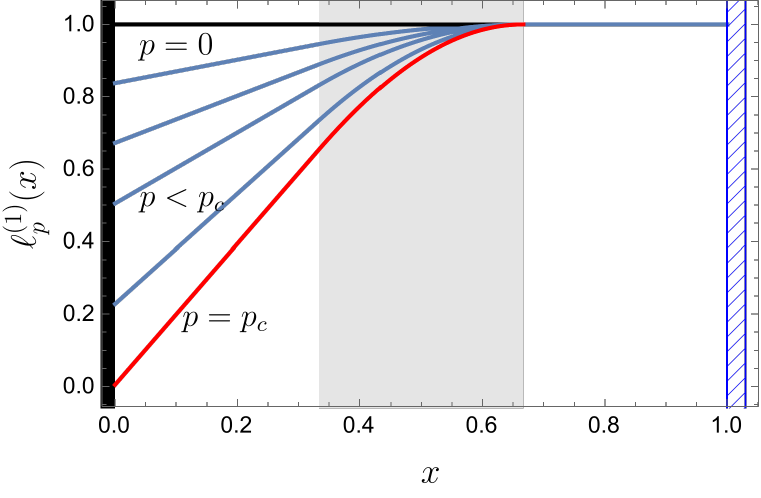}
    \caption{(color online) The left eigenfunction $\ell^{(1)}_p(x)$ in \eqref{eq:fp B abs} corresponding to the vanishing eigenvalue $\mu(p)=0$ for different values of $p$ and a fixed set of parameters $a=1/3$, $b=2/3$, $L=1$. {The curves from top to bottom correspond to values of $p=0$, $1$, $2$, $3$, $5$, and $6.67~(p_c)$.} At the critical value $p=p_c$, the eigenfunction $\ell^{(1)}_p(x)$, indicated in red, vanishes at the absorbing boundary $x=0$.}
    \label{fig:eigfn_nonquantum}
\end{figure}
\begin{figure}
    \centering
    \includegraphics[width=0.48 \textwidth]{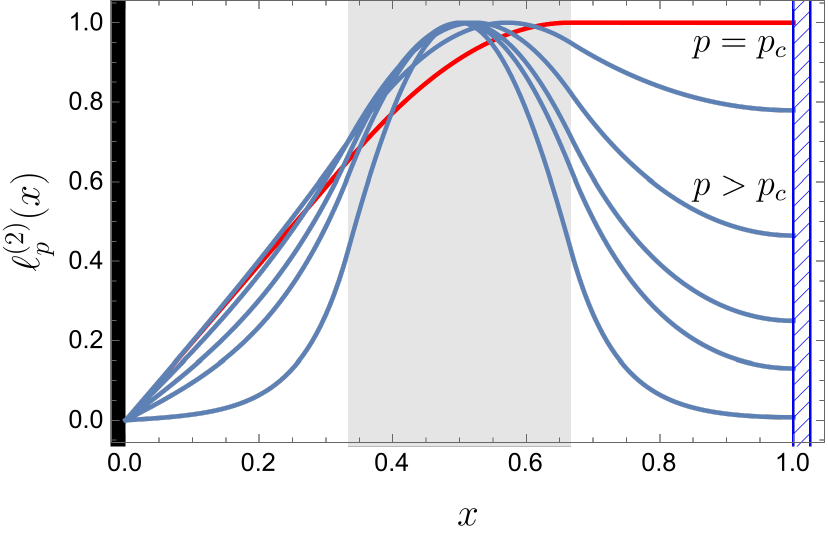}
    \caption{(color online) The left eigenfunction $\ell^{(2)}_p(x)$ in \eqref{eq:lp BM abs} with the absorbing boundary condition at $x=0$ for identical set of parameter values of $a$, $b$, and $L$ as in \fref{fig:eigfn_nonquantum}. {The curves from top to bottom at the right boundary correspond to $p=6.67~(p_c)$, $15$, $30$, $50$, $75$, and $250$, respectively.} At the critical value $p=p_c$, the eigenfunction $\ell_p(x)$, indicated in red, is constant for $x>b$. }
    \label{fig:eigfn_quantum}
\end{figure}

\subsubsection{Zero eigenvalue problem}
The left eigenfunction $\ell^{(1)}_p(x)$ corresponding to the zero eigenvalue $\mu_1(p)=0$ with the reflecting boundary condition $\ell^{(1)\prime}_p(L)=0$ is a piece-wise function
\begin{equation}
    \ell^{(1)}_p(x)= \begin{cases}  A_1+A_2 x & \textrm{for }0<x<a \\ B_1\sin(\omega x)+B_2 \cos(\omega x) & \textrm{for } x \in[a, b] \\ 1 & \textrm{for } b<x<L \end{cases},\label{eq:fp B abs}
\end{equation} 
where $\omega^2=p$ and $A_1, A_2, B_1$, and $B_2$ are $p$-dependent constants determined by continuity of $\ell^{(1)}_p(x)$ and its derivative. The eigenfunction is plotted in \fref{fig:eigfn_nonquantum} for a set of values of the tilting parameter $p$. Notably, $\ell^{(1)}_0(x)=1$, which is the left eigenfunction associated with the stationary state of a Markov operator. It is worth mentioning that the eigenfunction does not necessarily vanish at the absorbing wall $x=0$.

\subsubsection{Eigenvalue-problem with absorbing boundary condition}\label{sec:absorbingBC}
The eigenvalue problem for $\mu_2(p)$ and its associated left eigenfunction $\ell^{(2)}_p(x)$, with the absorbing boundary condition $\ell^{(2)}_p(0)=0$ and reflecting boundary condition $\ell_p^{(2)\prime}(L)=0$, is analogous to the quantum mechanical problem with infinite potential at origin, a square-well potential in $a<x<b$, and a zero current condition at $x=L$. The solution is given by
\begin{equation}
    \ell^{(2)}_p(x)= \begin{cases}  e^{\kappa x}-e^{-\kappa x} & 0<x<a \\ D_1 \sin(\gamma x) +D_2 \cos (\gamma x) & x \in[a, b] \\ E \left(e^{\kappa (x-L)}+e^{-\kappa (x-L)}\right) & b<x<L \end{cases},\label{eq:lp BM abs}
\end{equation}
where $\kappa=\sqrt{ \mu_2(p)}$,  $\gamma=\sqrt{p-\mu_2(p)}$, and the $p$-dependent coefficients $D_1$, $D_2$ and $E$ are determined from continuity of the eigenfunction and its derivative. The eigenvalue $\mu_2(p)$ is a solution of the transcendental equation
\begin{equation}
\frac{v+u \tan (v (b-a)) \tanh (u (L-b))}{v \{v \tan (v (b-a))-u \tanh (u (L-b))\}}=\frac{\tanh (a u)}{u}\label{eq:transcedental abs spectral}
\end{equation}
where $u^2=\mu_2(p)$ and $v^2=p-\mu_2(p)$. The eigenfunction $\ell_p^{(2)}(x)$ is illustrated in \fref{fig:eigfn_quantum} for a range of tilting parameters $p$ above a critical value $p_c$ defined by the solution of \eqref{eq:transcedental abs spectral} for $\mu_2(p_c)=0$.

As in the previous example, the crossing of the two eigenvalues, $\mu_1(p)=0$ and $\mu_2(p)$, gives rise to a non-analytic scgf, which in turn results in a non-analytic ldf. These findings match with those obtained in \sref{sec:BM abs res laplace}.

\subsubsection{Conditioned dynamics}

\begin{figure}
    \centering
    \includegraphics[width=0.48 \textwidth]{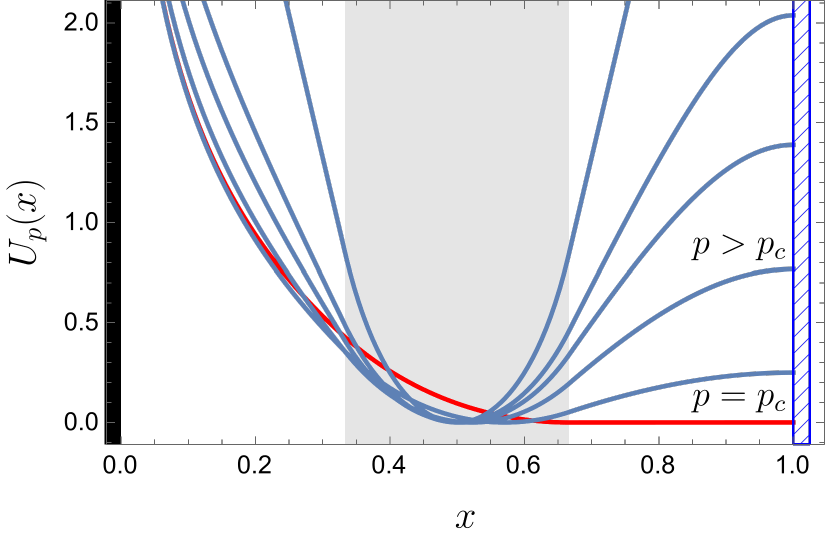}
    \caption{(color online) The effective potential corresponding to the eigenfunction $\ell_p^{(2)}(x)$ for the dynamics of a Brownian particle confined between an absorbing wall at the origin and a reflecting wall at $x=1$, conditioned to yield a value of residence time $Q=T\mu'(p)$. The plot corresponds parameter values $a=1/3, b=2/3$, and $L=1$. {The curves from bottom to top correspond to increasing values of $p=6.67~(p_c)$, $15$, $30$, $50$, $75$, and $250$. The effective potential becomes progressively steeper as $p$ increases.}}
    \label{fig:potential_nonquantum}
\end{figure}

The singularity of the ldf can be attributed to abrupt change in the conditioned dynamics. Referring back to the discussion at the beginning of \sref{sec:tiltedintro} and \eqref{eq:effective continume Langevin}, the restricted ensemble of trajectories spending a fraction of time $q=\mu'(p)$ in the region $[a,b]$, can be described as a Brownian motion inside a potential
\begin{equation}
    U_p(x)=- 2 \ln \ell_p(x).\label{eq:effective U BM abs 2}
\end{equation}
For $p>p_c$, where the scgf is $\mu_2(p)$, the eigenfunction $\ell_p^{(2)}(x)$ gives a potential (see \fref{fig:potential_nonquantum}) that diverges at $x=0$ and consequently, the trajectories never reach the absorbing site. As $q\to 1$, $\ell_p^{(2)}(x)\to 0$ for $x$ outside $[a,b]$ resulting in a confining potential $U_p(x)$ around the interval $[a,b]$.

For $p<p_c$, where the scgf $\mu(p)=0$, the relation $q=\mu'(p)$ yields a constant value for $q$. This indicates a breakdown of ensemble equivalence, rendering the construction \cite{derrida2019large} of effective dynamics inapplicable. Further discussions on this can be found in \cite{nyawo2018dynamical}. Nevertheless, it is evident that the conditioned dynamics for $p>p_c$ do not extend to $p<p_c$, indicating a sudden change of the dynamics at the transition point $p=p_c$.

\subsection{Residence time for mortal Brownian motion}
\label{sec:sticky tilted}


We now return to the problem discussed in \sref{sec:stickyresidencelaplace}: Brownian particle subjected to a constant death rate, with the observable being the total time spent within an interval $x\in[-a,a]$. Once again, the scgf, $\mu(p)$, of the residence time is given by the largest eigenvalue of the corresponding tilted operator.

Adhering to the framework elucidated in \cite{derrida2019large}, it can be shown that the pertinent tilted operator for this problem takes the form of 
\begin{equation}
    \mathcal{L}_p= \begin{pmatrix}
    0 & \alpha \\
    0 & \widehat{\mathcal{L}}_p
    \end{pmatrix}.\label{eq:L op MB}
\end{equation}
where 
\begin{equation}
\widehat{\mathcal{L}}_p=\frac{d^2}{dx^2} -\alpha  +p\mathbb{1}_{[-a,a]}.
    \label{eq:tilted L sticky}
\end{equation}
In writing $\mathcal{L}_p$, we separated the configuration space based on whether the particle is alive or dead. The dead particle is immobile and does not contribute to our observables. Consequently, a dead particle is assigned a single configuration by ignoring its position degrees of freedom.

The scgf, $\mu(p)$, for the residence time is given by the largest eigenvalue of $\mathcal{L}_p$. In this context, there are two distinct relevant eigenvalues. One of them is $\mu_1(p)=0$ for all $p$ associated to the left eigenvector
$\begin{pmatrix}
    1 & \ell^{(1)}_p(x)
    \end{pmatrix}$,
with
\begin{equation}
    \ell^{(1)}_p(x)= \begin{dcases}  1+A e^{\sqrt{\alpha} x} & \textrm{for }x<-a \\ B\cosh(x\sqrt{\alpha-p} )+\frac{\alpha}{\alpha+p} & \textrm{for } x \in[-a, a] \\
    1+A e^{-\sqrt{\alpha} x} & \textrm{for } x>a, \end{dcases}
    \label{eqn:fun_fp}
\end{equation}
where the constants $A$ and $B$ are determined by the continuity of $\ell^{(1)}_p(x)$ and its derivative. It is noteworthy that, for $p=0$, $\ell^{(1)}_0(x)=1$ which is consistent with the unit left eigenvector corresponding to zero eigenvalue of a Markov operator. 

\begin{figure}
    \centering
    \includegraphics[width=0.48 \textwidth]{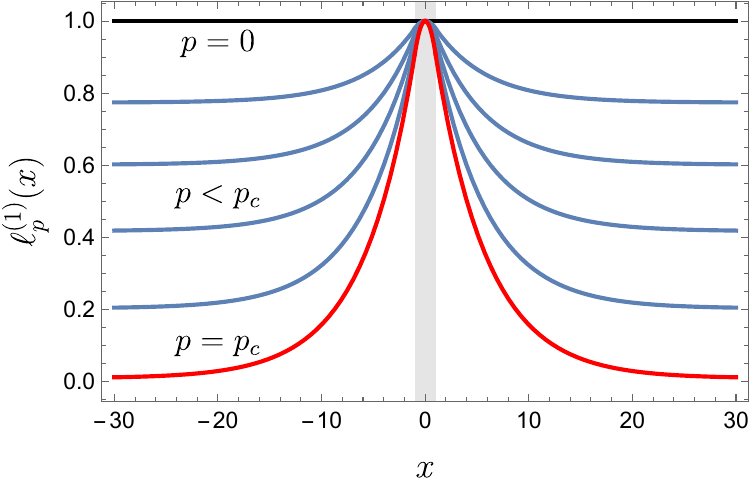}
    \caption{(color online) The normalized functional component of the eigenvector $\begin{pmatrix}
1 & \ell^{(1)}_p(x)
\end{pmatrix}$ corresponding to the eigenvalue $\mu(p)=0$ of the operator $\mathcal{L}_p$ is depicted for various values of $p<p_c$. 
{The curves from top to bottom correspond to increasing values of $p=0$, $0.094$, $0.126$, $0.158$, $0.194$, and $0.23~(p_c)$. The function exhibits progressively steeper decay from its maximum as $p$ increases.}  For $p=p_c$, highlighted in red, the function asymptotically vanishes for $x\pm \infty$. The normalization is such that the maximum value is set to $1$. The parameters used are $a=1$ and $\alpha=0.04$, while the shaded gray region indicates the interval $-1<x<1$.}
\label{fig:eigfn_sticky_nonquantum}
\end{figure}
\begin{figure}
    \centering
    \includegraphics[width=0.48 \textwidth]{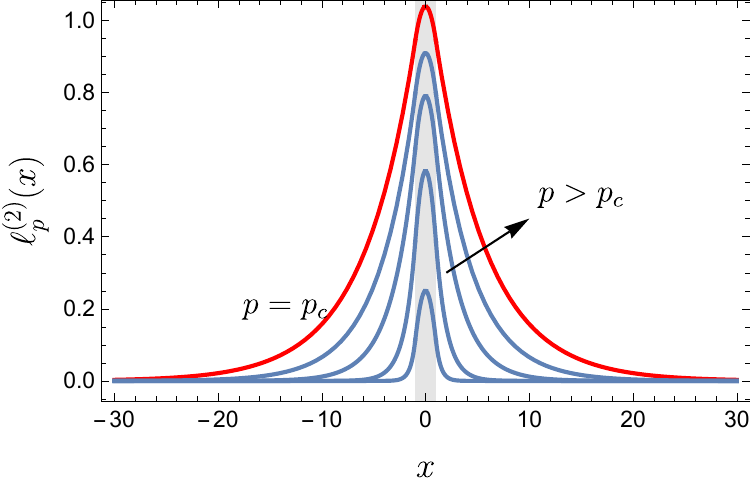}
    \caption{(color online) The function $\ell_p^{(2)}(x)$ in \eqref{eqn:eigfn_sticky} is a component of the eigenvector $\begin{pmatrix}
    0 & \ell^{(2)}_p(x)
    \end{pmatrix}$ associated with the eigenvalue $\mu_2(p)$ in \eqref{eq:mu2p MBM} of the operator $\mathcal{L}_p$ in \eqref{eq:L op MB} for the same set of parameter values used in \fref{fig:eigfn_sticky_nonquantum}.
    {The curves from top to bottom correspond to increasing values of $p=0.23~(p_c)$, $0.33$, $0.58$, $1.62$, and $6$. The maximum of the function decreases with increasing $p$.} The red  line corresponds to $p=p_c$ while the gray region indicates the interval $-1<x<1$.}
    \label{fig:eigfn_sticky_quantum}
\end{figure}

The second eigenvalue, $\mu_2(p)$, corresponds to the eigenvector
$\begin{pmatrix}
    0 & \ell^{(2)}_p(x)
    \end{pmatrix}$,
where $\ell^{(2)}_p(x)$ is the left eigenfunction corresponding to the largest eigenvalue $\mu_2(p)$ of $\widehat{\mathcal{L}}_p$. The operator in \eqref{eq:tilted L sticky} is related by a similarity transformation to the tilted operator for residence time of a drifted BM \cite{nyawo2018dynamical}. It is also analogous to the quantum Hamiltonian for a particle in a finite potential well. This quantum analogy straightforwardly gives the largest eigenvalue of $\widehat{\mathcal{L}}_p$ written as
\begin{equation}
    \mu_2(p)=\lambda(p)-\alpha.\label{eq:mu2p MBM}
\end{equation}
where $\lambda(p)$ is the largest solution to the transcendental equation
\begin{equation}
    \zeta=\gamma\tan(\gamma a)
    \label{eqn:sticky_trans}
\end{equation}
with $\zeta=\sqrt{\lambda(p)}$ and $\gamma=\sqrt{p-\lambda(p)}$. The corresponding left eigenfunction is 
\begin{equation}
    \ell^{(2)}_p(x)= \begin{cases}  e^{\zeta x} & \textrm{for }x<-a \\ C\cos(\gamma x) & \textrm{for } x \in[-a, a] \\
    e^{-\zeta x} & \textrm{for } x>a. \end{cases}
    \label{eqn:eigfn_sticky}
\end{equation}
with the parameter $C$ determined from the continuity condition at $x=\pm a$.
\begin{figure}
    \centering
    \includegraphics[width=0.48 \textwidth]{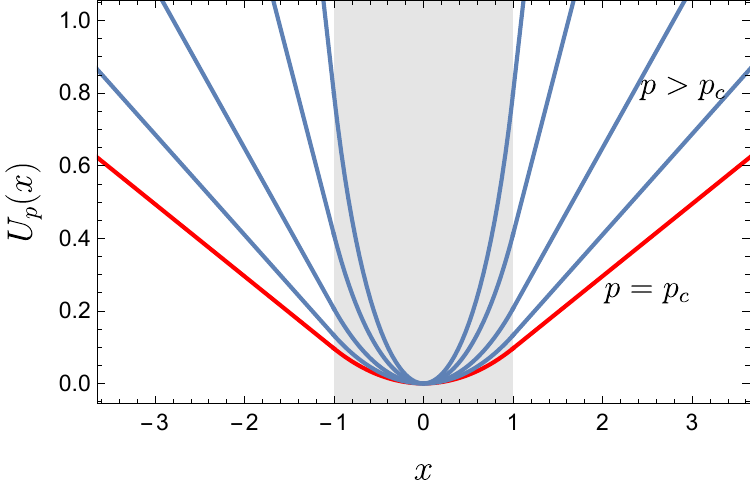}
    \caption{{(color online) The effective potential in \eqref{eq:effective U BM abs} for a mortal Brownian particle, conditioned to yield a value of residence time $Q=T\mu'(p)$. The plot corresponds to parameter values $a=1$, and $\alpha=0.04$. The curves from bottom to top correspond to increasing values of $p=0.23~(p_c)$, $0.33$, $0.58$, $1.62$, and $6$. The effective potential becomes progressively steeper as $p$ increases.}}
    \label{fig:potential_nonquantum_mortal}
\end{figure}

The scgf $\mu(p)$ for the residence time is the largest among the two eigenvalues: $\mu_1(p)=0$ and $\mu_2(p)$, which cross each other at a critical value $p=p_c$, reproducing the piece-wise function in \eqref{eq:cgf sticky residence} and consequently, the associated ldf in \eqref{eq:ldf sticky residence}. (We have explicitly verified the agreement with the result in \eqref{eq:cgf sticky residence}.) The value of $p_c$ can be determined by setting $\lambda(p_c)=\alpha$ in \eqref{eqn:sticky_trans}, for which $\mu_2(p_c)$ in \eqref{eq:mu2p MBM} becomes zero. 

The eigenvectors associated with the leading eigenvalues, $\mu_1(p)$ and $\mu_2(p)$, in their respective domains, are shown in \fref{fig:eigfn_sticky_nonquantum} and \fref{fig:eigfn_sticky_quantum}, respectively, for a range of tilting parameters $p$.

\subsubsection*{Conditioned dynamics}
As in the previous example, the singularity of the ldf here is due to the abrupt change in the conditioned dynamics. Beyond a critical value $p_c$, where the scgf is given by the eigenvalue $\mu_2(p)$, the effective mortality rate vanishes. This is seen from the vanishing component of the eigenvector $\begin{pmatrix}0 & \ell^{(2)}_p(x)\end{pmatrix}$ in the effective transition rates constructed by comparing $\mathcal{L}_p$ in \eqref{eq:L op MB} with the tilted matrix of a discrete-state Markov process discussed in \cite{derrida2019large}. As a result, the effective dynamics resembles that of a standard Brownian motion conditioned to spend $q=Q/T$ fraction of time in the interval $[-a,a]$, which, following \eqref{eq:effective continume Langevin}, is described by a Brownian motion inside an effective potential
\begin{equation}
    U_p(x)=-2 \ln \ell_p^{(2)}(x)\label{eq:effective U BM abs}
\end{equation}
with $\ell_p^{(2)}(x)$ in \eqref{eqn:eigfn_sticky}.
The bias parameter $p$ is related to the observable's value by the ensemble equivalence relation $q=\mu'(p)$. The vanishing of $\ell_p^{(2)}(x)$ for large $x$ (see Fig.~\ref{fig:eigfn_sticky_quantum}) results in a confining potential $U_p(x)$ around the interval $[-a,a]$, which gets narrower as $q\to 1$ (equivalently larger $p$). {The effective potential is shown in Fig.~\ref{fig:potential_nonquantum_mortal}.} For $p<p_c$, the ensemble equivalence breaks down and the construction of the effective dynamics in \cite{derrida2019large} does not apply.

\section{Markov Chains}\label{sec:markovchains}
The origin of DPTs, as discussed in our examples of Langevine processes, is relatively easier to comprehend in their discrete counterparts, which are discrete time Markov processes on a finite configuration space. For a Markov chain, evolution of the probability $P_t(C)$ of a configuration $C$ at time $t$ is governed by
\begin{equation}
P_{t+1}(C')=\sum_C M(C',C)P_t(C),
\end{equation}
where $M(C',C)$ denotes the transition probability from state $C$ to $C'$. A discrete analogue \cite{derrida2019large} of the empirical observable \eqref{eqn:residencetime} is  
\begin{equation}
    Q=\sum_{t=0}^{T-1} f(C_t)+\sum_{t=0}^{T-1} g(C_{t+1},C_t),
    \label{eqn:markovobservable}
\end{equation}
where $f(C)$ and $g(C,C')$ are arbitrary functions. For instance, the residence time at a configuration $C$ corresponds to $f(C_t)=\delta_{C_t,C}$ and $g(C_{t+1},C_t)=0$, while $f(C_t)=0$ and $g(C_{t+1},C_t)=\delta_{C_{t}, C}\delta_{C_{t+1}, C'}$ gives the total number of jumps from configuration $C$ to $C'$. 

In the limit of large $T$, the probability of $Q$ has the asymptotic \eqref{eqn:largedeviation} with the corresponding scgf $\mu(p)$ in \eqref{eqn:scgf} relating to the largest eigenvalue $e^{\mu(p)}$ of the tilted matrix \cite{jack2010large, jack2015effective,chetrite2013nonequilibrium,chetrite2015nonequilibrium,derrida2019large}:
\begin{equation}
    M_p(C,C')=M(C,C') e^{p[f(C')+g(C,C')]},
    \label{eqn:tiltedmatrix}
\end{equation}
which is a discrete analogue of the tilted operator \eqref{eqn:tilted}.

Analogous to \eqref{eq:effective continume Langevin}, the dynamics conditioned to yield a value of $Q_T$, in the large time limit, are described (excluding the non-quasi-stationary regimes near $t=0$ and $t=T$) by an effective Markov chain \cite{jack2010large, jack2015effective,chetrite2013nonequilibrium,chetrite2015nonequilibrium,derrida2019large}:
\begin{equation}
P_{t+1}^{(p)}(C')=\sum_C W_p(C',C)P_t^{(p)}(C),
\end{equation}
with the transition matrix 
\begin{equation}
    W_p(C',C)=\frac{L_p(C')}{e^{\mu(p)}L_p(C)}M_p(C',C),
    \label{eqn:tiltedmarkov}
\end{equation}
where $L_p$ is the left eigenvector of $M_p$ corresponding to the eigenvalue $e^{\mu(p)}$.
The tilting parameter $p$ is expressed in terms of the observable value $Q_T$ by the relation $\mu'(p)=Q_T/T$ \cite{jack2010large, jack2015effective,chetrite2013nonequilibrium,chetrite2015nonequilibrium,derrida2019large}. 

In this formalism, the crossing of two leading eigenvalues of $M_p$ results a singular scgf and, consequently, an abrupt change in the effective transition matrix $W_p$. This draws an instructive analogy between DPTs and conventional equilibrium phase transitions relating to crossing of two largest eigenvalues of transfer matrix \cite{cuesta2004general}. There are known examples of eigenvalue crossings in the transfer matrix of one-dimensional models \cite{cuesta2004general,kittel1969phase}. Notably, the transfer matrices in these examples are reducible and therefore do not satisfy the criteria for the Perron-Frobenius theorem \cite{cuesta2004general}. In the following examples, we shall illustrate that the DPTs in non-ergodic Markov chains share similar characteristics, arising from the reducibility of the corresponding tilted matrices \eqref{eqn:tiltedmatrix}. 

\subsection{Residence time in a three-state Markov chain}\label{sec:residencediscrete}

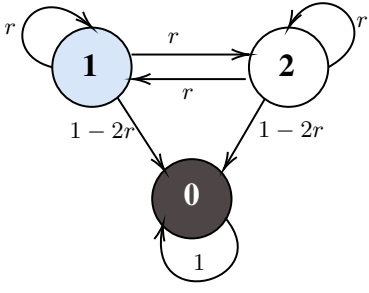
\begin{figure}
\centering
\begin{tikzpicture}[x=0.5pt,y=0.5pt,yscale=-1.2,xscale=1.2]

\draw  [fill={rgb, 255:red, 215; green, 230; blue, 248 }  ,fill opacity=1 ] (100,134) .. controls (100,120.19) and (111.19,109) .. (125,109) .. controls (138.81,109) and (150,120.19) .. (150,134) .. controls (150,147.81) and (138.81,159) .. (125,159) .. controls (111.19,159) and (100,147.81) .. (100,134) -- cycle ;
\draw   (224,134) .. controls (224,120.19) and (235.19,109) .. (249,109) .. controls (262.81,109) and (274,120.19) .. (274,134) .. controls (274,147.81) and (262.81,159) .. (249,159) .. controls (235.19,159) and (224,147.81) .. (224,134) -- cycle ;
\draw  [fill={rgb, 255:red, 78; green, 70; blue, 70 }  ,fill opacity=1 ] (163,217) .. controls (163,203.19) and (174.19,192) .. (188,192) .. controls (201.81,192) and (213,203.19) .. (213,217) .. controls (213,230.81) and (201.81,242) .. (188,242) .. controls (174.19,242) and (163,230.81) .. (163,217) -- cycle ;
\draw    (141,153) -- (169.92,197.82) ;
\draw [shift={(171,199.5)}, rotate = 237.17] [color={rgb, 255:red, 0; green, 0; blue, 0 }  ][line width=0.75]    (10.93,-3.29) .. controls (6.95,-1.4) and (3.31,-0.3) .. (0,0) .. controls (3.31,0.3) and (6.95,1.4) .. (10.93,3.29)   ;
\draw    (150,126) -- (222,126) ;
\draw [shift={(224,126)}, rotate = 180] [color={rgb, 255:red, 0; green, 0; blue, 0 }  ][line width=0.75]    (10.93,-3.29) .. controls (6.95,-1.4) and (3.31,-0.3) .. (0,0) .. controls (3.31,0.3) and (6.95,1.4) .. (10.93,3.29)   ;
\draw    (234,154) -- (209.01,196.77) ;
\draw [shift={(208,198.5)}, rotate = 300.3] [color={rgb, 255:red, 0; green, 0; blue, 0 }  ][line width=0.75]    (10.93,-3.29) .. controls (6.95,-1.4) and (3.31,-0.3) .. (0,0) .. controls (3.31,0.3) and (6.95,1.4) .. (10.93,3.29)   ;
\draw    (100,134) .. controls (62.38,124.6) and (86.51,73.54) .. (123.86,107.93) ;
\draw [shift={(125,109)}, rotate = 223.85] [color={rgb, 255:red, 0; green, 0; blue, 0 }  ][line width=0.75]    (10.93,-3.29) .. controls (6.95,-1.4) and (3.31,-0.3) .. (0,0) .. controls (3.31,0.3) and (6.95,1.4) .. (10.93,3.29)   ;
\draw    (274,134) .. controls (314.39,111.84) and (270.36,72.7) .. (249.92,107.36) ;
\draw [shift={(249,109)}, rotate = 298.07] [color={rgb, 255:red, 0; green, 0; blue, 0 }  ][line width=0.75]    (10.93,-3.29) .. controls (6.95,-1.4) and (3.31,-0.3) .. (0,0) .. controls (3.31,0.3) and (6.95,1.4) .. (10.93,3.29)   ;
\draw    (209,229) .. controls (242.66,269.1) and (157.73,292.04) .. (169.61,234.28) ;
\draw [shift={(170,232.5)}, rotate = 103.13] [color={rgb, 255:red, 0; green, 0; blue, 0 }  ][line width=0.75]    (10.93,-3.29) .. controls (6.95,-1.4) and (3.31,-0.3) .. (0,0) .. controls (3.31,0.3) and (6.95,1.4) .. (10.93,3.29)   ;
\draw    (222,141.5) -- (153,141.5) ;
\draw [shift={(151,141.5)}, rotate = 360] [color={rgb, 255:red, 0; green, 0; blue, 0 }  ][line width=0.75]    (10.93,-3.29) .. controls (6.95,-1.4) and (3.31,-0.3) .. (0,0) .. controls (3.31,0.3) and (6.95,1.4) .. (10.93,3.29)   ;

\draw (117,124) node [anchor=north west][inner sep=0.75pt]   [align=left] {{\large {\fontfamily{ptm}\selectfont \textbf{1}}}};
\draw (241,124) node [anchor=north west][inner sep=0.75pt]   [align=left] {{\large {\fontfamily{ptm}\selectfont \textbf{2}}}};
\draw (181,205) node [anchor=north west][inner sep=0.75pt] [color={rgb, 255:red, 255; green, 255; blue, 255 }  ,opacity=1 ]  [align=left] {{\large {\fontfamily{ptm}\selectfont \textbf{0}}}};
\draw (171,110.4) node [anchor=north west][inner sep=0.75pt]    {$r$};
\draw (228,166.4) node [anchor=north west][inner sep=0.75pt]    {$1-2r$};
\draw (187,250.4) node [anchor=north west][inner sep=0.75pt]    {$1$};
\draw (180,145.4) node [anchor=north west][inner sep=0.75pt]    {$r$};
\draw (68,104.4) node [anchor=north west][inner sep=0.75pt]    {$r$};
\draw (290,100.4) node [anchor=north west][inner sep=0.75pt]    {$r$};
\draw (109,167.4) node [anchor=north west][inner sep=0.75pt]    {$1-2r$};

\end{tikzpicture}

\caption{(color online) A three-state Markov chain with one absorbing state denoted by $\textbf{0}$. The arrows indicate the allowed transitions with the corresponding transition probabilities. The residence time is measured at the state $\textbf{1}$ (shaded in blue).}
\label{fig:sitesthree}
\end{figure}

An example of a three state Markov chain is shown in \fref{fig:sitesthree}. The corresponding transition matrix is
\begin{equation}
    M= \begin{pmatrix}
    1 & 1-2r & 1-2r \\
    0 & r & r \\
    0 & r & r \\
    \end{pmatrix}.
\end{equation}
The observable we consider is the residence time in the state $\textbf{1}$, which corresponds to $f(C)=\delta_{C,\textbf{1}}$ and $g=0$ in the definition \eqref{eqn:markovobservable}.

The tilted matrix \eqref{eqn:tiltedmatrix} for this observable
\begin{equation}
    M_p= \begin{pmatrix}
    1 & (1-2r) e^p & 1-2r \\
    0 & r e^p & r \\
    0 & r e^p & r \\
    \end{pmatrix}.
\end{equation}
has three eigenvalues: $e_1=1$, $e_2=r(1+e^p)$, and $e_3=0$. The left eigenvectors of the dominant eigenvalues $e_1$ and $e_2$ are 
\begin{equation}
v_1= \begin{pmatrix}
    \dfrac{1-r-r e^p}{1-2r} & e^p & 1
\end{pmatrix}
\label{eqn:eigvecdiscretetime}
\end{equation}
and $v_2= \begin{pmatrix}
    0 & e^p & 1
\end{pmatrix}$,
respectively.

The scgf is given by the logarithm of the largest among the three eigenvalues, leading to
\begin{align}
\mu(p)&=\max\{0,\ln (r(1+e^p))\}
\label{eqn:scgfmax}
\end{align}
and its Legendre-Fenchel transformation \eqref{eqn:legendre} gives the piece-wise ldf
\begin{align}
\phi(q)&=
\begin{dcases}
q \ln\left(\frac{1-r}{r}\right) &\textrm{for }0\leq q \leq q^\star\\
q \ln\left(\frac{q}{1-q}\right)-\ln\left(\frac{r}{1-q}\right) & \textrm{for } q^\star < q \leq 1
\end{dcases}
\label{eqn:ratefndiscreteresidence}
\end{align}
where $q^\star=1-r$. 

The shape of this ldf is qualitatively similar to the ldf in Fig \ref{fig:wall_time_ratefn} for the residence time of the Brownian motion in presence of an absorbing site.

\begin{figure}

\begin{tikzpicture}[x=0.5pt,y=0.5pt,yscale=-1.2,xscale=1.2]

\draw  [fill={rgb, 255:red, 215; green, 230; blue, 248 }  ,fill opacity=1 ] (100,134) .. controls (100,120.19) and (111.19,109) .. (125,109) .. controls (138.81,109) and (150,120.19) .. (150,134) .. controls (150,147.81) and (138.81,159) .. (125,159) .. controls (111.19,159) and (100,147.81) .. (100,134) -- cycle ;
\draw   (224,134) .. controls (224,120.19) and (235.19,109) .. (249,109) .. controls (262.81,109) and (274,120.19) .. (274,134) .. controls (274,147.81) and (262.81,159) .. (249,159) .. controls (235.19,159) and (224,147.81) .. (224,134) -- cycle ;
\draw  [fill={rgb, 255:red, 78; green, 70; blue, 70 }  ,fill opacity=1 ] (163,217) .. controls (163,203.19) and (174.19,192) .. (188,192) .. controls (201.81,192) and (213,203.19) .. (213,217) .. controls (213,230.81) and (201.81,242) .. (188,242) .. controls (174.19,242) and (163,230.81) .. (163,217) -- cycle ;
\draw    (150,126) -- (222,126) ;
\draw [shift={(224,126)}, rotate = 180] [color={rgb, 255:red, 0; green, 0; blue, 0 }  ][line width=0.75]    (10.93,-3.29) .. controls (6.95,-1.4) and (3.31,-0.3) .. (0,0) .. controls (3.31,0.3) and (6.95,1.4) .. (10.93,3.29)   ;
\draw    (100,134) .. controls (62.38,124.6) and (86.51,73.54) .. (123.86,107.93) ;
\draw [shift={(125,109)}, rotate = 223.85] [color={rgb, 255:red, 0; green, 0; blue, 0 }  ][line width=0.75]    (10.93,-3.29) .. controls (6.95,-1.4) and (3.31,-0.3) .. (0,0) .. controls (3.31,0.3) and (6.95,1.4) .. (10.93,3.29)   ;
\draw    (274,134) .. controls (314.39,111.84) and (270.36,72.7) .. (249.92,107.36) ;
\draw [shift={(249,109)}, rotate = 298.07] [color={rgb, 255:red, 0; green, 0; blue, 0 }  ][line width=0.75]    (10.93,-3.29) .. controls (6.95,-1.4) and (3.31,-0.3) .. (0,0) .. controls (3.31,0.3) and (6.95,1.4) .. (10.93,3.29)   ;
\draw    (222,141.5) -- (153,141.5) ;
\draw [shift={(151,141.5)}, rotate = 360] [color={rgb, 255:red, 0; green, 0; blue, 0 }  ][line width=0.75]    (10.93,-3.29) .. controls (6.95,-1.4) and (3.31,-0.3) .. (0,0) .. controls (3.31,0.3) and (6.95,1.4) .. (10.93,3.29)   ;

\draw (117,124) node [anchor=north west][inner sep=0.75pt]   [align=left] {{\large {\fontfamily{ptm}\selectfont \textbf{1}}}};
\draw (241,124) node [anchor=north west][inner sep=0.75pt]   [align=left] {{\large {\fontfamily{ptm}\selectfont \textbf{2}}}};
\draw (181,205) node [anchor=north west][inner sep=0.75pt] [color={rgb, 255:red, 255; green, 255; blue, 255 }  ,opacity=1 ]  [align=left] {{\large {\fontfamily{ptm}\selectfont \textbf{0}}}};
\draw (170,91.4) node [anchor=north west][inner sep=0.75pt]    {\large{$\frac{1}{e^p+1}$}};
\draw (170,143.4) node [anchor=north west][inner sep=0.75pt]    {\large{$\frac{e^p}{e^p+1}$}};
\draw (40,104.4) node [anchor=north west][inner sep=0.75pt]    {\large{$\frac{e^p}{e^p+1}$}};
\draw (290,100.4) node [anchor=north west][inner sep=0.75pt]    {\large{$\frac{1}{e^p+1}$}};
\draw (253,214.4) node [anchor=north west][inner sep=0.75pt]    {$p \geq p_{c}$};

\end{tikzpicture}
 \caption{(color online) Effective Markov processes for $p>p_c$ obtained using \eqref{eqn:tiltedmarkov}. In this scenario, the Markov process effectively behaves as if the absorbing site is not present, and the Markov trajectories do not evolve into it.\label{fig:effective rates three state}}
\end{figure}
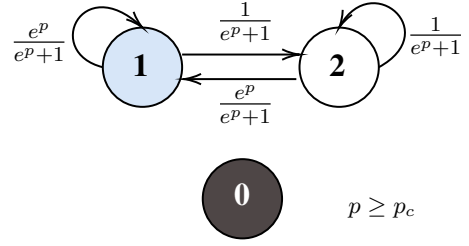

The transition rates of the effective Markov process \eqref{eqn:tiltedmarkov} for tilting parameter $p>p_c$ are expressed in terms of the eigenvector of the dominant eigenvalue and sketched in \fref{fig:effective rates three state}. However, for $p<p_c$, ensemble equivalence breaks down, rendering the construction of effective dynamics in \cite{derrida2019large} inapplicable.

{ \textit{Remark:} The ldf in \eqref{eqn:ratefndiscreteresidence} can alternatively be obtained using combinatorics. There are two kinds of trajectories contributing to the probability $P_T(Q)$ of $Q$ at time $T$: those which survive until time $T$ from reaching the absorbing site $0$ (alive) and those which do not (dead). Assuming that the process starts at the site $2$, we get 
\begin{align}
    P_T\left(Q\right)&=P_T^{(a)}(Q)+P_T^{(d)}(Q) \\    
    &=\binom{T}{Q}r^{T}+(1-2r)\sum_{n=0}^{T-Q-1}\binom{Q+n}{n}r^{Q+n}. \nonumber
    \label{eqn:discreteresidencepdf}
\end{align}
Stirling's approximation and a saddle point calculation for large $T$ recover the asymptotic \eqref{eqn:largedeviation} with the ldf given in \eqref{eqn:ratefndiscreteresidence}. }

\textit{Remark:} The two-state version of this problem was studied in \cite{coghi2019large}, where, although the scgf was non-analytic, the ldf $\phi(q)$ was found to be linear across the entire allowed range of $q$.

\subsection{Integrated current in a four-state Markov chain}
\label{sec:discretevelocity} 
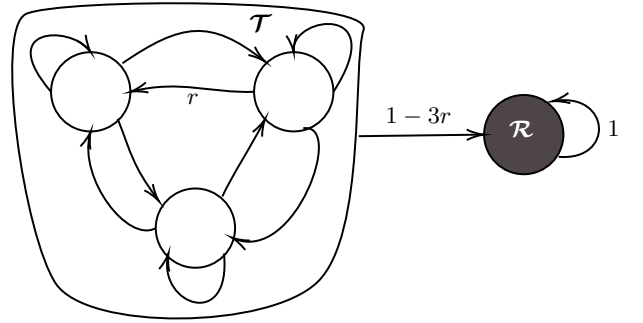
\begin{figure}
    \centering
\begin{tikzpicture}[x=0.6pt,y=0.6pt,yscale=-1,xscale=1]

\draw   (116,76) .. controls (116,62.19) and (127.19,51) .. (141,51) .. controls (154.81,51) and (166,62.19) .. (166,76) .. controls (166,89.81) and (154.81,101) .. (141,101) .. controls (127.19,101) and (116,89.81) .. (116,76) -- cycle ;
\draw   (244,76) .. controls (244,62.19) and (255.19,51) .. (269,51) .. controls (282.81,51) and (294,62.19) .. (294,76) .. controls (294,89.81) and (282.81,101) .. (269,101) .. controls (255.19,101) and (244,89.81) .. (244,76) -- cycle ;
\draw   (182,162) .. controls (182,148.19) and (193.19,137) .. (207,137) .. controls (220.81,137) and (232,148.19) .. (232,162) .. controls (232,175.81) and (220.81,187) .. (207,187) .. controls (193.19,187) and (182,175.81) .. (182,162) -- cycle ;
\draw  [fill={rgb, 255:red, 78; green, 70; blue, 70 }  ,fill opacity=1 ] (389,103) .. controls (389,89.19) and (400.19,78) .. (414,78) .. controls (427.81,78) and (439,89.19) .. (439,103) .. controls (439,116.81) and (427.81,128) .. (414,128) .. controls (400.19,128) and (389,116.81) .. (389,103) -- cycle ;
\draw   (100,27) .. controls (120,17) and (320,16) .. (313,37) .. controls (306,58) and (315,166) .. (284,196) .. controls (253,226) and (139,227) .. (119,197) .. controls (99,167) and (80,37) .. (100,27) -- cycle ;
\draw    (161,58) .. controls (198.43,32.39) and (216.46,30.07) .. (249.48,55.8) ;
\draw [shift={(251,57)}, rotate = 218.45] [color={rgb, 255:red, 0; green, 0; blue, 0 }  ][line width=0.75]    (10.93,-3.29) .. controls (6.95,-1.4) and (3.31,-0.3) .. (0,0) .. controls (3.31,0.3) and (6.95,1.4) .. (10.93,3.29)   ;
\draw    (275,99) .. controls (301.6,95.06) and (269.98,184.26) .. (231.75,168.78) ;
\draw [shift={(230,168)}, rotate = 25.97] [color={rgb, 255:red, 0; green, 0; blue, 0 }  ][line width=0.75]    (10.93,-3.29) .. controls (6.95,-1.4) and (3.31,-0.3) .. (0,0) .. controls (3.31,0.3) and (6.95,1.4) .. (10.93,3.29)   ;
\draw    (182,162) .. controls (164.36,171.8) and (133.27,130.7) .. (140.51,102.7) ;
\draw [shift={(141,101)}, rotate = 107.82] [color={rgb, 255:red, 0; green, 0; blue, 0 }  ][line width=0.75]    (10.93,-3.29) .. controls (6.95,-1.4) and (3.31,-0.3) .. (0,0) .. controls (3.31,0.3) and (6.95,1.4) .. (10.93,3.29)   ;
\draw    (244,76) .. controls (216.56,76.98) and (197.76,68.36) .. (167.85,75.54) ;
\draw [shift={(166,76)}, rotate = 345.53] [color={rgb, 255:red, 0; green, 0; blue, 0 }  ][line width=0.75]    (10.93,-3.29) .. controls (6.95,-1.4) and (3.31,-0.3) .. (0,0) .. controls (3.31,0.3) and (6.95,1.4) .. (10.93,3.29)   ;
\draw    (158,93) .. controls (166.82,115.54) and (167,121.75) .. (183,143.64) ;
\draw [shift={(184,145)}, rotate = 233.53] [color={rgb, 255:red, 0; green, 0; blue, 0 }  ][line width=0.75]    (10.93,-3.29) .. controls (6.95,-1.4) and (3.31,-0.3) .. (0,0) .. controls (3.31,0.3) and (6.95,1.4) .. (10.93,3.29)   ;
\draw    (224,142) .. controls (230.79,127.45) and (241.34,111.96) .. (250.18,94.61) ;
\draw [shift={(251,93)}, rotate = 116.57] [color={rgb, 255:red, 0; green, 0; blue, 0 }  ][line width=0.75]    (10.93,-3.29) .. controls (6.95,-1.4) and (3.31,-0.3) .. (0,0) .. controls (3.31,0.3) and (6.95,1.4) .. (10.93,3.29)   ;
\draw    (116,76) .. controls (81.08,35.26) and (127.08,34.04) .. (139.91,49.52) ;
\draw [shift={(141,51)}, rotate = 237.09] [color={rgb, 255:red, 0; green, 0; blue, 0 }  ][line width=0.75]    (10.93,-3.29) .. controls (6.95,-1.4) and (3.31,-0.3) .. (0,0) .. controls (3.31,0.3) and (6.95,1.4) .. (10.93,3.29)   ;
\draw    (294,76) .. controls (330.08,20.42) and (269.19,27.59) .. (268.9,49.3) ;
\draw [shift={(269,51)}, rotate = 262.57] [color={rgb, 255:red, 0; green, 0; blue, 0 }  ][line width=0.75]    (10.93,-3.29) .. controls (6.95,-1.4) and (3.31,-0.3) .. (0,0) .. controls (3.31,0.3) and (6.95,1.4) .. (10.93,3.29)   ;
\draw    (225,178) .. controls (229.9,227) and (183.9,210.69) .. (189.59,179.9) ;
\draw [shift={(190,178)}, rotate = 104.04] [color={rgb, 255:red, 0; green, 0; blue, 0 }  ][line width=0.75]    (10.93,-3.29) .. controls (6.95,-1.4) and (3.31,-0.3) .. (0,0) .. controls (3.31,0.3) and (6.95,1.4) .. (10.93,3.29)   ;
\draw    (310,104) -- (387,103.03) ;
\draw [shift={(389,103)}, rotate = 179.27] [color={rgb, 255:red, 0; green, 0; blue, 0 }  ][line width=0.75]    (10.93,-3.29) .. controls (6.95,-1.4) and (3.31,-0.3) .. (0,0) .. controls (3.31,0.3) and (6.95,1.4) .. (10.93,3.29)   ;
\draw    (436,117) .. controls (471.46,119.96) and (469.08,76.34) .. (433.64,81.72) ;
\draw [shift={(432,82)}, rotate = 349.29] [color={rgb, 255:red, 0; green, 0; blue, 0 }  ][line width=0.75]    (10.93,-3.29) .. controls (6.95,-1.4) and (3.31,-0.3) .. (0,0) .. controls (3.31,0.3) and (6.95,1.4) .. (10.93,3.29)   ;

\draw (239,25.4) node [anchor=north west][inner sep=0.75pt]    {$\mathbf{\mathbfcal{T}}$};
\draw (403,92.4) node [anchor=north west][inner sep=0.75pt]  [color={rgb, 255:red, 255; green, 255; blue, 255 }  ,opacity=1 ]  {$\mathbf{\mathbfcal{R}}$};
\draw (325,84.4) node [anchor=north west][inner sep=0.75pt]    {$1-3r$};
\draw (200,75.4) node [anchor=north west][inner sep=0.75pt]    {$r$};
\draw (465,92.4) node [anchor=north west][inner sep=0.75pt]    {$1$};

\end{tikzpicture}
    \caption{A four state Markov chain featuring a single recurrent state $\mathbfcal{R}$ and a transient state space $\mathbfcal{T}$ consisting of three states. In $\mathbfcal{T}$, any transient state has a probability $1-3r$, with $r<\frac{1}{3}$, to escape to $\mathbfcal{R}$. For jumps within $\mathbfcal{T}$ the transition probability is $r$ as indicated by arrows.}
    \label{fig:currentdiscrete}
\end{figure}

The Markov chain in consideration is illustrated in \fref{fig:currentdiscrete}. This example serves a simple discrete analogue of the mortal Brownian motion considered in \ref{sec:stickybrownian}, where the recurrent state is equivalent to the immobile dead state of the Brownian. 

\begin{figure}
    \centering   \includegraphics[width=0.47\textwidth]{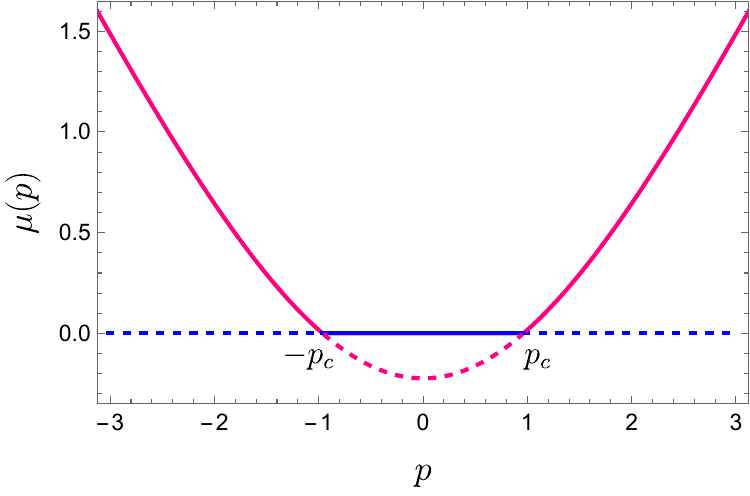}
    \caption{(color online) The solid line, obtained by the maximum of the two largest eigenvalues \eqref{eqn:scgfmax2}, represents the scgf for the integrated current in the four-state Markov chain in \fref{fig:currentdiscrete} with parameter value $r=0.2$.}
    \label{fig:discrete_current_2_scgf}
\end{figure}

The observable we consider is the integrated current $Q$ in $T$-time steps counted by adding $1$ for each clockwise jump and subtracting $1$ for an anti-clockwise jump within $\mathcal{T}$. Jumps to $\mathcal{R}$ or to remain at the same site do not contribute to $Q$.  The tilted matrix \eqref{eqn:tiltedmatrix} for this observable is
\begin{equation}
    M_{p}=\begin{pmatrix}
    
    1 & 1-3r & 1-3r & 1-3r \\
    0  & r & r \mathrm{e}^{-p} & r \mathrm{e}^p \\
    0  & r \mathrm{e}^p & r & r \mathrm{e}^{-p} \\
    0 & r \mathrm{e}^{-p} & r \mathrm{e}^p & r
    \end{pmatrix},
\end{equation}
From the largest eigenvalue of $M_p$ the scgf is 
\begin{equation}
\mu(p)=\max\{0,e_2(p)\}
\label{eqn:scgfmax2}
\end{equation}
where $e_2(p)=\ln(r+2r\cosh{p})$. The Legendre-Fenchel transformation \eqref{eqn:legendre} of the scgf gives the piece-wise ldf
\begin{equation}
\phi(q)=
\begin{dcases}
 p_c |q| & |q| \leq q_c\\
f(|q|) & |q| > q_c,
\end{dcases}
\label{eqn:ratemax}
\end{equation}
where $p_c=\arccosh\frac{(1-r)}{2r}$, $q_c=\sqrt{(1-3 r) (1+r)}$ and 
\begin{equation}
\begin{split}
    f(q)= q \log \left(\frac{u}{2(1-q)}\right)-\log \left(\left(\frac{1}{1-q}+\frac{4}{u}\right) r\right)
\end{split}
\end{equation}
with $u=\sqrt{4-3 q^2}+q$.
The scgf \eqref{eqn:scgfmax2} plotted in \fref{fig:discrete_current_2_scgf} shows two singularities. Corresponding ldf is similar to the ldf of position (equivalent of integrated current $\int dt \dot{X}_t$ ) in  \fref{fig:one_sticky} for the mortal Brownian particle. 


\section{A robust mechanism for DPTs}\label{sec:generalidea}
A common feature between the two Markov chain examples in \sref{sec:markovchains} is the presence of an absorbing state, from which the system, once entered, can not escape. Their Markov matrices and the associated tilted matrices are reducible making them beyond the purview of the Perron-Forbenius theorem \cite{cuesta2004general}. As a result, crossing of largest eigenvalues is permitted, leading to a singularity in the scgf and, equivalently, a DPT.

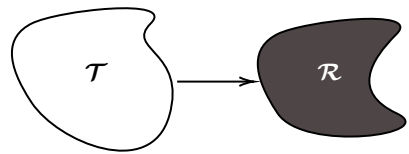
\begin{figure}
\centering
\begin{tikzpicture}[x=0.5pt,y=0.5pt,yscale=-1,xscale=1]

\draw  [fill={rgb, 255:red, 78; green, 70; blue, 70 }  ,fill opacity=1 ] (272,80) .. controls (292,70) and (382,60) .. (362,80) .. controls (342,100) and (328,124) .. (362,140) .. controls (396,156) and (292,170) .. (272,140) .. controls (252,110) and (252,90) .. (272,80) -- cycle ;
\draw   (86,72) .. controls (106,62) and (196,52) .. (176,72) .. controls (156,92) and (204,93) .. (190,142) .. controls (176,191) and (106,162) .. (86,132) .. controls (66,102) and (66,82) .. (86,72) -- cycle ;
\draw    (196,116) -- (252,116) ;
\draw [shift={(254,116)}, rotate = 180] [color={rgb, 255:red, 0; green, 0; blue, 0 }  ][line width=0.75]    (10.93,-3.29) .. controls (6.95,-1.4) and (3.31,-0.3) .. (0,0) .. controls (3.31,0.3) and (6.95,1.4) .. (10.93,3.29)   ;

\draw (124,100) node [anchor=north west][inner sep=0.75pt]   [align=left] {$\mathbf{\mathbfcal{T}}$};
\draw (300,100) node [anchor=north west][inner sep=0.75pt] [color={rgb, 255:red, 255; green, 255; blue, 255 }  ,opacity=1 ]  [align=left] {$\mathbf{\mathbfcal{R}}$};

\end{tikzpicture}
\caption{Schematic representation of Markov processes featuring a transient state space $\mathcal{T}$ where the system, once leaving $\mathcal{T}$, cannot reenter from the subspace $\mathcal{R}$.}
\label{fig:sitesgeneral}
\end{figure}

This mechanism for DPTs is expected in general for processes with transient states (see illustration in \fref{fig:sitesgeneral}). The transient state space $\mathcal{T}$ is not accessible from the recurrent state space  $\mathcal{R}$. Corresponding Markov matrix has the form:
\begin{equation}
    \mathbf{M}\equiv \begin{pmatrix}
    \mathbf{\mathbfcal{T}} & \mathbf{0}\\
    \mathbfcal{I} & \mathbf{\mathbfcal{R}}
    \end{pmatrix},
    \label{eqn:generalmatrix}
\end{equation}
where $\mathbfcal{T}$ and  $\mathbfcal{R}$ are blocks representing transitions within the sub-spaces $\mathcal{T}$ and $\mathcal{R}$, respectively, while $\mathbfcal{I}$ involves transitions from $\mathcal{T}$ to $\mathcal{R}$. 

The set of eigenvalues of $\mathbf{M}$ is composed of eigenvalues from $\mathbfcal{T}$ and $\mathbfcal{R}$. At large times, the transient state space $\mathcal{T}$ becomes empty, and the system reaches the stationary state in $\mathcal{R}$. The stationary distribution is given by the eigenvector of $\mathbfcal{R}$ with the largest eigenvalue $e_1=1$. This implies that the largest eigenvalue (denoted as $e_2$) of $\mathbfcal{T}$ is less than $1$.

For an empirical observable measured in the transient state (as was the case for the examples discussed), only the block $\mathbfcal{T}$ is weighted by the tilting parameter $p$. This means that the spectrum of $\mathbfcal{R}$ remains unchanged with $e_1=1$ for all $p$. However, for large $p$, it is expected that the scgf, which is logarithm of the largest eigenvalue of the tilted matrix $\mathbf{M}_p$, is positive, implying that $e_2>1$ for large $p$. This would occur if $e_2$, which was smaller than $e_1=1$ for $p=0$, crossed $e_1$ at some intermediate value of $p$, resulting in a DPT.

\textit{Remark:} The transient states for the examples discussed in this article are explicit. In the example of a drifted Brownian motion, discussed in \cite{nyawo2018dynamical}, any finite position is transient as the particle eventually drifts away to infinity.

\section{Multiple phase transitions}\label{sec:multipledpts}
Building upon the mechanism of DPT discussed in this article, it is possible to construct simple examples with rich phase behaviours. In the following sections, we explore two such simple examples.

\subsection{Epidemiological process}\label{sec:evaporation}
Consider a grim epidemiological process where infected individuals quickly die without any chance of recovery. The death rate for an individual is proportional to the number of living persons $n(t)$ at that instant $t$. The scenario is analogous to a variant of Verhulst model \cite{Meerson_2013} of population dynamics with unit reproduction rate. The average number of living persons $\bar{n}(t)$ at a time follows $\frac{d\bar{n}(t)}{dt}\simeq -\alpha\, \bar{n}(t)^2$ with $\alpha$ being a parameter, leading to an algebraic decay $\bar{n}(t)\simeq N/(1+\alpha t N)$, where $N$ is the initial size of the population. 

\begin{figure}
    \centering
    \includegraphics[width=0.5 \textwidth]{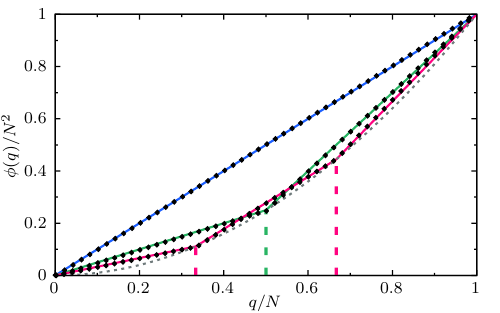}
    \caption{(color online) The ldf \eqref{eq:ldf eva exact} for a set of initial population size $N$ (color code: blue $=1$, green $=2$, red $=3$). In the large $N$ limit, the scaled ldf $\phi(q)/N^2$ converges to the dotted curve $x^2$. The data points indicate results from importance sampling simulations for $T = 100$ with $dt = 0.01$ (see Appendix.~\ref{app:importance sampling}).}
    \label{fig:multi_evaporating_ratefn}
\end{figure}

A relevant observable is the net resources consumed by the living persons over a period $T$, which is, at the zeroth level approximation and up to a proportionality constant, 
\begin{equation}
    Q=\int_0^T n(t)dt.
    \label{eqn:evapobservable}
\end{equation}
{On an average the consumed resources increase slowly in time $T$ as $\bar{Q}=\alpha^{-1}\log (1+\alpha N T)$ while the fluctuations follow the large deviation }asymptotics \eqref{eqn:largedeviation}, featuring a piece-wise ldf (see Appendix \ref{sec:multievaporate_derivation} for a derivation)
\begin{equation}
    \phi(q)=q(2k+1)-k(k+1)\quad \textrm{for $k \leq q < k+1$} \label{eq:ldf eva exact}
\end{equation}
with $k=0,1,\ldots, N-1$.  The multiple derivative discontinuities of the ldf are shown in \fref{fig:multi_evaporating_ratefn}. These discontinuities correspond to first order DPTs between phases characterized by the maximum number of surviving persons $k$ at the time $T$.


\subsection{Mortal Brownians}
\label{sec:multisticky}
A simple extension of the previous example is where the mortal particles diffuse in space. This is a multi-particle generalization of the mortal Brownians introduced in \sref{sec:stickybrownian}, where death rate of an individual is proportional to the number of living.   

As an observable, we consider the net displacement of all particles
\begin{equation}
    Q=\int_0^T \sum_{i=1}^N dx_{i}(t)\label{eq:multiple sicky q}
\end{equation}
where $x_i(t)$ is the position of the $i$-th particle at time $t$, and $N$ is the size of the population (living and dead together).

The probability of $Q$ for large time $T$ follows the asymptotics \eqref{eqn:largedeviation} with a piece-wise ldf composed of alternating domains where $\phi(q)$ is linear and quadratic (see \fref{fig:multi_evaporating_ratefn}).  At the boundaries of these domains $\phi''(q)$ is discontinuous, indicating a second-order DPT between phases characterized by the number of surviving particles at time $T$. We defer expression of $\phi(q)$ and its derivation in the Appendix \ref{sec:stickyderivation}.
\begin{figure}
    \centering
    \includegraphics[width=0.5 \textwidth]{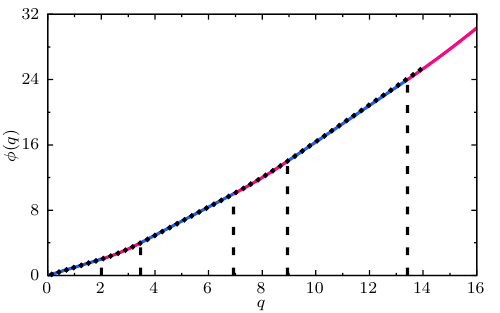}
    \caption{(color online) The ldf $\phi(q)$ for the probability of $Q=qT$ in \eqref{eq:multiple sicky q} for three mortal Brownians. Only positive $q$ results are shown. The red color indicates quadratic dependence, while the blue color represents linear dependence. The data points indicate results from importance sampling simulations for $T = 100$ with $dt = 0.01$ (see Appendix.~\ref{app:importance sampling}).}
    \label{fig:three_sticky}
\end{figure}



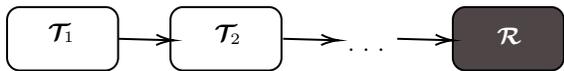
\begin{figure}
\centering

\begin{tikzpicture}[x=0.6pt,y=0.6pt,yscale=-1,xscale=1]

\draw   (120,133) .. controls (120,128.58) and (123.58,125) .. (128,125) -- (182,125) .. controls (186.42,125) and (190,128.58) .. (190,133) -- (190,157) .. controls (190,161.42) and (186.42,165) .. (182,165) -- (128,165) .. controls (123.58,165) and (120,161.42) .. (120,157) -- cycle ;
\draw   (223,133) .. controls (223,128.58) and (226.58,125) .. (231,125) -- (285,125) .. controls (289.42,125) and (293,128.58) .. (293,133) -- (293,157) .. controls (293,161.42) and (289.42,165) .. (285,165) -- (231,165) .. controls (226.58,165) and (223,161.42) .. (223,157) -- cycle ;
\draw  [fill={rgb, 255:red, 78; green, 70; blue, 70 }  ,fill opacity=1 ] (401,133) .. controls (401,128.58) and (404.58,125) .. (409,125) -- (463,125) .. controls (467.42,125) and (471,128.58) .. (471,133) -- (471,157) .. controls (471,161.42) and (467.42,165) .. (463,165) -- (409,165) .. controls (404.58,165) and (401,161.42) .. (401,157) -- cycle ;
\draw    (190,145) -- (223,145.94) ;
\draw [shift={(225,146)}, rotate = 181.64] [color={rgb, 255:red, 0; green, 0; blue, 0 }  ][line width=0.75]    (10.93,-3.29) .. controls (6.95,-1.4) and (3.31,-0.3) .. (0,0) .. controls (3.31,0.3) and (6.95,1.4) .. (10.93,3.29)   ;
\draw    (294,145) -- (327,145.94) ;
\draw [shift={(329,146)}, rotate = 181.64] [color={rgb, 255:red, 0; green, 0; blue, 0 }  ][line width=0.75]    (10.93,-3.29) .. controls (6.95,-1.4) and (3.31,-0.3) .. (0,0) .. controls (3.31,0.3) and (6.95,1.4) .. (10.93,3.29)   ;
\draw    (366,145) -- (399,145.94) ;
\draw [shift={(401,146)}, rotate = 181.64] [color={rgb, 255:red, 0; green, 0; blue, 0 }  ][line width=0.75]    (10.93,-3.29) .. controls (6.95,-1.4) and (3.31,-0.3) .. (0,0) .. controls (3.31,0.3) and (6.95,1.4) .. (10.93,3.29)   ;

\draw (333,149.4) node [anchor=north west][inner sep=0.75pt]    {$.\ .\ .$};
\draw (144,133.4) node [anchor=north west][inner sep=0.75pt]    {$\mathbfcal{T}_{1}$};
\draw (249,134.4) node [anchor=north west][inner sep=0.75pt]    {$\mathbfcal{T}_{2}$};
\draw (427,136.4) node [anchor=north west][inner sep=0.75pt]  [color={rgb, 255:red, 255; green, 255; blue, 255 }  ,opacity=1 ]    {$\mathbfcal{R}$};

\end{tikzpicture}

\caption{A Markov process with a sequence of transient state spaces $\mathcal{T}_i$ such that evolution can take place only in the direction indicated destined to a recurrent state.}
\label{fig:multipleDPTs}
\end{figure}

\textit{Remark:} The multiple sequence of DPTs in these two examples relate to a transient structure of the state space illustrated in Fig \ref{fig:multipleDPTs}. As particles get absorbed, the system moves from one transient state space to the next. A similar sequence of multiple DPTs was recently observed in the context of the occupation fraction of vicious walkers \cite{mukherjee2023dynamical}.

\section{Non-Markov processes}\label{sec:nonmarkovextend}
The mechanism of DPT that we discussed in \sref{sec:generalidea} extends for non-Markovian processes. The key idea is that a non-Markov process with finite memory can be described by a higher dimensional Markov process, which also poses transient sectors if the original process have them. One would expect DPTs for such processes. We give two such examples.


\subsection{Mortal Active Ornstein-Uhlenbeck process}
The active Ornstein-Uhlenbeck process \cite{martin2021statistical} is a simple model of persistent dynamics for a non-polar active particle. The position of the particle follows a Langevin dynamics
\begin{equation}
    \ddot{X}_t+\frac{1}{\tau}\dot{X}_t=\eta_t 
\end{equation}
with a Gaussian white noise $\eta_t$ with zero-mean and covariance $\langle \eta_t\eta_{t'}\rangle=2\delta(t-t')$. The non-zero auto-correlation $\langle \dot{X}_t\dot{X}_{t'}\rangle \sim \tau e^{-\vert t -t'\vert/\tau}$ for large $t-t'$ makes the $X_t$ dynamics non-Markovian, although on the $(X_t,\dot{X}_t)$ space the process is Markovian.

The free propagator for the position of AOUP started at the origin $X_0=0$ with $\dot{X}_0=0$ is Gaussian:
\begin{equation}
g_t(x)= \frac{1}{\sqrt{4\pi t\mathcal{D}_t}}\exp\left(-\frac{x^2}{4t\mathcal{D}_t}\right),\label{eq:prop aoup}
\end{equation}
with the variance
\begin{equation}
    \mathcal{D}_t=\tau ^2 \left(1-\frac{\tau  }{2 t}\left(3-4 e^{-\frac{t}{\tau }}+e^{-\frac{2 t}{\tau }}\right)\right).
\end{equation}

For a mortal AOUP with a death rate $\alpha$, the probability of the particle position $x$ at time $t$ can be written as \eqref{eqn: stickydistribution} with the propagator $g_t(x)$ given in \eqref{eq:prop aoup}. For large $t$, the probability has the asymptotic $P_t(x=qt)\sim e^{-t \phi(q/\tau)}$ with the ldf that is identical to \eqref{eq:phi sticky 1}.

This coincidence of the ldf with the Brownian case is not surprising considering that the measurement time $t $ is far larger than the persistance time $\tau$ for the AOUP. 

\subsection{Mortal fractional Brownian motion (fBm)}\label{sec:fBm}
An fBm \cite{sadhu2021functionals,sadhu2018generalized} is a non-Markovian Gaussian process $X_t$ with mean zero and power-law auto-correlation:
\begin{equation}
    \langle \dot{X}_t \dot{X}_{t'}\rangle =2H(2H-1) (t -t')^{2(H-1)}
\end{equation}
for $t>t'>0$ with the Hurst exponent $0<H<1$. The free propagator for a fractional Brownian starting at the origin is given by the Gaussian 
\begin{equation}
g_t(x)= \frac{1}{\sqrt{4\pi t^{2H}}}\exp\left(-\frac{x^2}{4t^{2H}}\right).\label{eq:prop fbm}
\end{equation}

For a mortal fBm with a death rate $\alpha$, the probability of position at time $t$ is similar to \eqref{eqn: stickydistribution} with the propagator in \eqref{eq:prop fbm}. In the limit of large $t$, the probability has the large deviation asymptotic $P_t(x=q t^\nu)\sim e^{-t \phi(q)}$ with the ldf
\begin{subequations}

\begin{equation}
    \phi(q)=\begin{dcases}
    \beta \;|q|^{\frac{1}{\nu}} & \text{for} \ |q| \leq \sqrt{2\alpha/H} \\
    \alpha+\frac{q^2}{4} & \text{for} \ |q|> \sqrt{2\alpha/H}     
    \end{dcases}
\end{equation}
where $\nu=H+1/2$ and 
\begin{equation}
    \beta=\frac{\nu \alpha}{H}\left(\frac{H}{2\alpha}\right)^{\frac{1}{2\nu}}.
\end{equation}
\end{subequations}

\textit{Remark:} The same ldf was found for an fBm with resetting \cite{majumdar2015dynamical}.

\section{Conclusions} \label{sec:conclusions}
We constructed various illustrative simple models demonstrating DPTs in the statistics of time-integrated observables. These DPTs correspond to sudden changes in effective dynamics that correspond to fluctuations of the observables. For a mathematical origin, we have shown how such DPTs generically arise from the crossing of eigenvalues in reducible Markov operators for stochastic dynamics with transient states. Understanding this scenario aided us in constructing rich phase behaviors consisting of multiple DPTs and extensions for non-Markov processes. 

Extending this work for interacting many-body systems and their higher-dimensional generalizations presents intriguing possibilities. For instance, one could explore microscopic models of the epidemiological process in \sref{sec:evaporation} and population dynamics \cite{OVASKAINEN_2010,Ottino_2020}, where the density-dependent mortality rate emerges from interactions. Another avenue for investigation could involve studying fluctuations of empirical observables such as integrated current or entropy production in many-particle dynamics with absorbing states. Relying on the mechanism discussed in this work, it is reasonable to predict DPTs in such many-body generalizations. A well-known example \cite{di2023current} is the dynamical phase transition in the current fluctuations of resetting Brownian particles starting from a domain wall initial state on an infinite line. Similar transitions are anticipated for the many-particle generalization of mortal Brownian particles. For such extended systems, the framework of fluctuating hydrodynamics of conditioned dynamics \cite{Derrida2019} would provide an appropriate theoretical foundation.

In the mechanism discussed here, non-ergodicity arising from the underlying transient state-space is responsible for DPTs. Non-ergodicity may also effectively emerge in the thermodynamic limit of many-body systems or in the low-noise limit of single-degrees of freedom.  Exploring the nature and universality of DPTs in such non-ergodic dynamics warrants further investigation.

It is worth noting that that non-ergodicity is not a necessary criteria for DPTs. Singular ldfs are also observed in position distributions in run-and-tumble dynamics devoid of transient states. These singularities originate from a competition between trajectories with typical jumps and trajectories with a dominant atypical jump \cite{mori2021first,mori2021condensation}. Generalization of DPTs in such democratic versus winner-takes-all scenarios would be an interesting future direction. 

\section*{Acknowledgements}
YRY thanks Ritam Basu for computer access, Jagannath Rana for stimulating discussions, and Aravind Sugunan for assistance with computational techniques. SNM acknowledges the support from the Science and
Engineering Research Board (SERB, Government of India) under the VAJRA faculty scheme (No. VJR/2017/000110) and the  
ANR Grant No. ANR-23-CE30-0020-01 EDIPS. SNM and TS
thank the support from the International
Research Project (IRP) titled ‘Classical and quantum dynamics in out of equilibrium systems’ by CNRS, France. TS gratefully acknowledges CNRS for funding the visit to Laboratoire Charles Coulomb, Université de Montpellier, which initiated the part of the work on importance sampling.

\appendix
\section*{Appendices}
\renewcommand{\thesubsection}{\Alph{subsection}}

\subsection{Computer simulations \label{app:importance sampling}}
{ 
In this section, we present essential details of the importance-sampling techniques employed to numerically compute the large-deviation functions shown in Figs.~(\ref{fig:one_sticky},\ref{fig:one_sticky area},\ref{fig:sticky_residence_ratefn},\ref{fig:wall_local_time_ratefn},\ref{fig:wall_time_ratefn},\ref{fig:multi_evaporating_ratefn}) and Fig.~\ref{fig:three_sticky}.

\subsubsection{Computer simulations of the mortal Brownian walker}
For the simulations, we set $\alpha=1$ and $D=1$, which amounts to defining our units of time and length. At each time step, the walker dies with probability $p=dt$ and stays alive otherwise. In the limit $dt\to0$, this implies that the walker's lifetime follows an exponential distribution with parameter $\alpha$. 
For the simulations, $dt$ must be finite but small, and we have found that $dt=0.01$ is sufficient to capture the death process (we have verified that the first four cumulants of the walker's lifetime coincide with those calculated from the exponential distribution to within 1\%). Simulating one history of the mortal Brownian requires at most $M=\lfloor T/dt\rfloor$ pairs $\bm{\xi}=\lbrace (\nu_i,\eta_i)\rbrace_{i=1\dots M}$ of random numbers: $M$ random numbers $\lbrace \nu_i\rbrace_{i=1\dots M}$ drawn uniformly in $[0,\,1[$, and $M$ Gaussian random numbers $\lbrace \eta_i\rbrace_{i=1\dots M}$ of zero mean and variance $2dt$. Indeed, at each timestep $i$, the walker dies if $\nu_i<\alpha dt$, otherwise its position is shifted by an amount $\eta_i$.
\subsubsection{Computer simulations of multiple mortal Brownian walkers}
To simulate $N$ Brownian walkers with a death rate proportional to the number of alive walkers, we proceed as for one single walker. Computing one history now requires at most a set $\bm{\xi}$ of $M=\lfloor T/dt\rfloor$ $2N$-tuples $(\nu_{i,1},\dots \nu_{i,N},\eta_{i,1},\dots,\eta_{i,N})$, with $\nu_{i,n}\sim U(0,1)$ and $\eta_{i,n}\sim\mathcal{N}(0,2dt)$. At each timestep $i$, the $n$th walker dies if $\nu_{i,n}<N_{\mathrm{a}}dt$, with $N_{\mathrm{a}}$ the number of walkers which are still alive, otherwise its position is shifted by an amount $\eta_{i,n}$.

\subsubsection{Computer simulations of the epidemiological process}
The simulations of the epidemiological process are similar to the simulations of the multiple mortal Brownian walkers except that the particles do not diffuse. We set $\alpha=1$ which amounts to defining our unit of time. For the simulations, $dt$ must be finite but small, and we take $dt=0.01$. This value of $dt$ is small enough to measure accurately the average of $Q$ in (84) for times $t$ of order $1$, which reads 
\begin{equation}
    \langle Q\rangle =\begin{cases}
        1- e^{-t} \ \text{if} \ N=1,\\
        \dfrac{9-8e^{-t}-e^{-4t}}{6} \ \text{if} \ N=2,\\
        \dfrac{55-45e^{-t}-9e^{-4t}-e^{-9t}}{30} \ \text{if} \ N=3.\\
    \end{cases}
\end{equation}
Computing one history of the epidemiological process requires at most a set $\bm{\xi}$ of $M=\lfloor T/dt\rfloor$ $N$-tuples $(\nu_{i,1},\dots,\nu_{i,N})$ of random numbers drawn from a uniform distribution between 0 and 1.

\subsubsection{Computer simulations of the absorbing Brownian walker}
For the simulations, we set $L=1$ and $D=1$, which amounts to defining our units of length and time. At step $i=1\dots M$, the walker jumps from its current position $x_{i-1}$ to position $x_i=x_{i-1}+\eta_i$ with $\eta_i\sim\mathcal{N}(0,2dt)$. If $x_i>L$, then the particle is reflected at the boundary $x=L$, namely, $x_i\to 2L-x_i$. Instead, if $x_i<0$, then the particle is absorbed and the simulation ends. However, because of time discretization, the probability of absorption for $x=0$ is underestimated. In other words, because of the time discretization, one could have $x_{i-1}>0$ and $x_i>0$ while the trajectory may have crossed $x=0$ in the interval. The probability that such an event occurs can be computed analytically, and reads $e^{-x_{i-1}x_i/dt}$. As a consequence, simulating one history of the absorbing Brownian requires at most $M=\lfloor T/dt\rfloor$ pairs $\bm{\xi}=\lbrace (\eta_i,\nu_i)\rbrace_{i=1\dots M}$ of random numbers: $M$ Gaussian random numbers $\lbrace \eta_i\rbrace_{i=1\dots M}$ of zero mean and variance $2dt$, and $M$ random numbers $\nu_i$ drawn uniformly in $[0,\,1[$. Indeed, at each timestep $i$, the walker is absorbed if $\nu_i<\min(e^{-x_{i-1}x_i/dt},1)$~\cite{borodin2012handbook}. With this algorithm, we have found that $dt=10^{-3}$ allows us to simulate accurately the walker, in particular to capture the survival probability $S(t)$ and the distribution of the final position of the walker $P_t(x)$ (if it is still alive) on timescales $t$ of order 1, which are given by
\begin{equation}
    S(t)=\frac{4}{\pi}\sum_{k=0}^{+\infty}\frac{e^{-(\pi/2+k\pi)^2t}}{2k+1}\sin\left[\left(\frac{\pi}{2}+k\pi\right)x_0\right],
\end{equation}
with $x_0$ the initial position of the walker, and
\begin{align}
    P_t(x)=\sum_{k=0}^{+\infty}&e^{-(\pi/2+k\pi)^2t}\left\{\cos\left[\left(\frac{\pi}{2}+k\pi\right)(x-x_0)\right]\right.\nonumber\\
    &\left.-\cos\left[\left(\frac{\pi}{2}+k\pi\right)(x+x_0)\right]\right\}.
\end{align}
For the residence time, we set $x_0=1/2$ and measure the residence time in the interval $[1/3,\,2/3]$. For the local time we set $x_0=3/4$ and $a=1/2$. 

Measurements of the local time require extra random numbers. Because of the time discretization, the contribution to the local time is underestimated. To estimate it correctly, we use the known distribution of the contribution $dQ$ to the local time between steps $i-1$ (where the walker is at position $x_{i-1}$) and $i$ (where the walker is at position $x_i$)~\cite{borodin2012handbook}:
\begin{align}
    &P(dQ\vert x_i)=(1-p_{cr})\delta(dQ)+\frac{2dQ+|a-x_i|+|a-x_{i-1}|}{dt}\nonumber\\
    &\exp\left\{-\frac{(|a-x_i|+|a-x_{i-1}|+2dQ)^2-(x_i-x_{i-1})^2}{4dt}\right\},
\end{align}
where $p_{cr}=\min[e^{-(x_{i-1}-a)(x_i-a)/dt},1]$. The first term comes from the case where the walker has not crossed $x=a$, with $p_{cr}$ the probability that the path of the walker between steps $i-1$ and $i$ has crossed $x=a$, while the second term comes from the case where the walker has crossed $x=a$. As a consequence, with probability $1-p_{cr}$, $dQ=0$, while with probability $p_{cr}$, we have $dQ>0$, and its cumulative probability distribution reads
\begin{align}
    &P(dQ<z|dQ>0)=\frac{1}{p_{cr}}P(0<dQ<z)\\
    &=\frac{1}{p_{cr}}\int_0^{z}P(dQ\vert x_i)d(dQ)=1-e^{-\frac{z}{dt}(|a-x_i|+|a-x_{i-1}|+z)}.\nonumber
\end{align}
The distribution of $dQ$ knowing that $dQ>0$ can then easily be sampled by inverting the above cumulative probability distribution. We then conclude that 
\begin{align}
    dQ=\frac{1}{2}\bigg[&\sqrt{\left(|a-x_i|+|a-x_{i-1}|\right)^2-4dt\ln(1-\sigma)}\cr
    &-\left(|a-x_i|+|a-x_{i-1}|\right)\bigg],
\end{align}
where $\sigma\sim U(0,1)$. As a consequence, simulating one history of the absorbing Brownian to measure the distribution of the local time requires at most $M=\lfloor T/dt\rfloor$ 4-tuple $(\xi_i,\nu_i,\rho_i,\sigma_i)$ of random numbers, with $\xi_i\sim\mathcal{N}(0,2dt)$, and $\nu_i,\rho_i,\sigma_i\sim U(0,1)$. Indeed, at each timestep, the walker is absorbed if $\nu_i<\min(e^{-x_{i-1}x_i/dt},1)$. Besides, if the walker is not absorbed, it has crossed the level $x=a$ between steps $i-1$ and $i$ if $\rho_i<p_{cr}=\min[e^{-(x_{i-1}-a)(x_i-a)/dt},1]$. Finally, if the walker has crossed $x=a$ in the time interval, then the contribution $dQ_i$ to the local time at step $i$ is given by the above equation with $\sigma=\sigma_i$. The total local time then reads $Q=\sum_i dQ_i$.

\subsubsection{Importance sampling}

To sample the various large deviation functions, we use importance sampling techniques and adapt the procedure described in Ref.~\cite{hartmann2014high}. Any observable $q=Q/T^a$ (with $a$ depending on the observable) is a deterministic function $q=\hat q(\bm{\xi})$ of the random numbers $\bm{\xi}$ used to simulate an history. To sample the tails of the distribution of $q$, we implement a Markov Chain Monte Carlo on the histories, {\it i.e.}, to evolve the set $\bm{\xi}$ of random numbers~\cite{hartmann2024numerical}. More precisely, from the current set $\bm{\xi}$ of random numbers, corresponding to a value $\hat q(\bm{\xi})$ of the observable $q$, we propose to update the value of the random numbers for $M_{\mathrm{u}}$ timesteps randomly drawn from the entire set of $M$ pairs. From this trial set $\bm{\xi}_{\mathrm{trial}}$, we compute the trial value of the observable
$q_{\mathrm{trial}}=\hat q(\bm{\xi}_{\mathrm{trial}})$, and we accept this trial move with probability $\min(1,e^{-[W(q_{\mathrm{trial}})-W(q)]})$, with $W$ a weight function that depends only on the observable $q$ to sample. The value of $M_{\mathrm{u}}$ is chosen small enough to keep a reasonable acceptance rate of trial moves while ensuring an efficient sampling for all the weights considered below.

Due to our choice of Markov chain, the probability distribution of $q$ in the presence of the weight reads $P_T^{(W)}(q)=P_T(q)e^{-W(q)}/Z^{(W)}$, where $Z^{(W)}$ is a normalisation constant. In practice, we run $m$ simulations, each of them with a different harmonic weight $W_j(q)=\kappa(q-\bar{q}_j)^2$ (for $j=1\dots m$). The stiffness $\kappa$ is the same for all simulations, while we vary $\bar{q}_j$ from one simulation to another to explore rare fluctuations of $q$. For each simulation, we monitor the histogram $N_j(q)$, which counts the number of values of $\hat q(\bm{\xi})$ in the range $[q,\, q+dq)$. The values of the centres of the weights have been tuned in order to have a significant overlap between the histograms corresponding to adjacent values $\bar{q}_j$.

We then use the multi-histogram reweighting method~\cite{newman1999monte} to reconstruct $P_T(q)$ from the different histograms $N_j(q)$,
\begin{equation}
    P_T(q)dq = \frac{\sum_{j}N_j(q)}{\sum_jn_je^{-W_j(q)}/Z_j},
\end{equation}
with $n_j$ is the total number of values of $\hat q(\bm{\xi})$ measured, and where the partition functions $Z_j=Z^{(W_j)}$ are computed self-consistently from
\begin{equation}
    Z_j=\sum_q\frac{\sum_{j'}N_{j'}(q)}{\sum_{j'}n_{j'}e^{W_j(q)-W_{j'}(q)}/Z_{j'}}.
\end{equation}

To check that simulations with the harmonic bias have reached their steady state, we first run two simulations starting from different initial conditions, we monitor the time series of the observable $q$ and check that they converge to the same average value. We store the final sequence of random numbers. We then run longer simulations starting from the stored sequence of random numbers to build the histograms.

\subsubsection{Simulations parameters}
We present here the parameter values used to generate the importance sampling results shown in specific figures in the main text.

Figure~\ref{fig:one_sticky}. We have used $\kappa=100$, $M_{\mathrm{u}}=40$, and $m=35$ weights of 
centre $\bar{q}_j=-4.50$, $-4.25$, $-4.00$, $-3.75$, $-3.50$, $-3.25$, $-3.00$, $-2.75$, $-2.50$, $-2.25$, $-2.00$, $-1.75$, $-1.50$, $-1.25$, $-1.00$, $-0.75$, $-0.50$, $0.00$, $0.50$, $0.75$, $1.00$, $1.25$, $1.50$, $1.75$, $2.00$, $2.25$, $2.50$, $2.75$, $3.00$, $3.25$, $3.50$, $3.75$, $4.00$, $4.25$, $4.50$.

Figure~\ref{fig:one_sticky area}. We have used $\kappa=10000$, $M_{\mathrm{u}}=100$ and $m=136$ weights of centre $\bar{q}_j$ spaced by $0.02$ between $0.06$ and $1.40$ and between $-1.40$ and $-0.06$.\\

Figure~\ref{fig:sticky_residence_ratefn}. We have used $\kappa=10000$, $M_{\mathrm{u}}=40$ and $m=52$ weights of centre $\bar{q}_j$ equally spaced by $0.02$ between $-0.02$ and $1.00$.\\

Figure~\ref{fig:wall_local_time_ratefn}. We have used $\kappa=1000$, $M_{\mathrm{u}}=10$ and $m=57$ weights of centre $\bar{q}_j$ equally spaced by $0.05$ between $0.00$ and $2.80$.\\

Figure~\ref{fig:wall_time_ratefn}. We have used $\kappa=1000$, $m=31$ weights of centre $\bar{q}_j$ equally spaced by $0.05$ between $0.00$ and $1.50$, and $M_{\mathrm{u}}=30$ for $\bar{q}_j<1.15$ and $M_{\mathrm{u}}=10$ otherwise.\\

Figure~\ref{fig:multi_evaporating_ratefn}. We have used $\kappa=1000$ for all numbers of walkers $N=1,\,2,\,3$. For $N=1$ we have taken $M_{\mathrm{u}}=150$ while we have considered $M_{\mathrm{u}}=100$ for $N=2$ and 3. For $N=1$, we have used $m=20$ weights of centre $\bar{q}_j=0.00$, $0.10$, $0.15$, $0.20$, $0.25$, $0.30$, $0.35$, $0.40$, $0.45$, $0.50$, $0.55$, $0.60$, $0.65$, $0.70$, $0.75$, $0.80$, $0.85$, $0.90$, $0.95$, $1.00$. For $N=2$, we have used $m=58$ weights of centre $\bar{q}_j$ first equally spaced by $0.05$ between $0$ and $1.20$ and then spaced by $0.03$ between $1.20$ and $2.19$. For $N=3$, we have used $m=77$ weights of centre $\bar{q}_j$ first equally spaced by $0.10$ between $0.00$ and $0.20$, then equally spaced by $0.05$ between $0.20$ and $2.40$ and finally equally spaced by $0.03$ between $2.40$ and $3.30$.\\

Figure~\ref{fig:three_sticky}. We have used $\kappa=100$, $M_{\mathrm{u}}=40$, and $m=59$ weights of centre $\bar{q}_j$ first equally spaced by $0.50$ between $0.00$ and $0.50$, and then equally spaced by $0.25$ between $0.75$ and $14.75$.

}

\subsection{Boundary condition for absorbing wall} 
\label{sec:BCabsorbingappendix}

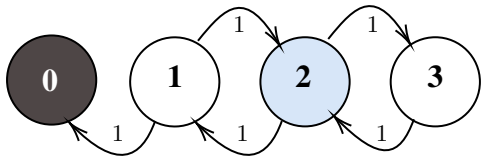
\begin{figure}
\centering
\begin{tikzpicture}[x=0.45pt,y=0.45pt,yscale=-1.5,xscale=1.5]

\draw   (141.02,136.99) .. controls (140.5,123.19) and (151.26,111.59) .. (165.06,111.06) .. controls (178.86,110.54) and (190.46,121.31) .. (190.98,135.1) .. controls (191.5,148.9) and (180.74,160.51) .. (166.94,161.03) .. controls (153.14,161.55) and (141.54,150.79) .. (141.02,136.99) -- cycle ;
\draw  [fill={rgb, 255:red, 215; green, 230; blue, 248 }  ,fill opacity=1 ] (214,136) .. controls (214,122.19) and (225.19,111) .. (239,111) .. controls (252.81,111) and (264,122.19) .. (264,136) .. controls (264,149.81) and (252.81,161) .. (239,161) .. controls (225.19,161) and (214,149.81) .. (214,136) -- cycle ;
\draw  [fill={rgb, 255:red, 78; green, 70; blue, 70 }  ,fill opacity=1 ] (72,135) .. controls (72,121.19) and (83.19,110) .. (97,110) .. controls (110.81,110) and (122,121.19) .. (122,135) .. controls (122,148.81) and (110.81,160) .. (97,160) .. controls (83.19,160) and (72,148.81) .. (72,135) -- cycle ;
\draw   (287,136) .. controls (287,122.19) and (298.19,111) .. (312,111) .. controls (325.81,111) and (337,122.19) .. (337,136) .. controls (337,149.81) and (325.81,161) .. (312,161) .. controls (298.19,161) and (287,149.81) .. (287,136) -- cycle ;
\draw    (179,113) .. controls (194.76,93.3) and (205.67,83.3) .. (224.15,114.54) ;
\draw [shift={(225,116)}, rotate = 240.07] [color={rgb, 255:red, 0; green, 0; blue, 0 }  ][line width=0.75]    (10.93,-3.29) .. controls (6.95,-1.4) and (3.31,-0.3) .. (0,0) .. controls (3.31,0.3) and (6.95,1.4) .. (10.93,3.29)   ;
\draw    (252,112) .. controls (267.76,92.3) and (278.67,82.3) .. (297.15,113.54) ;
\draw [shift={(298,115)}, rotate = 240.07] [color={rgb, 255:red, 0; green, 0; blue, 0 }  ][line width=0.75]    (10.93,-3.29) .. controls (6.95,-1.4) and (3.31,-0.3) .. (0,0) .. controls (3.31,0.3) and (6.95,1.4) .. (10.93,3.29)   ;
\draw    (226,158) .. controls (218.12,179.67) and (204.42,185.82) .. (179.16,160.19) ;
\draw [shift={(178,159)}, rotate = 46.08] [color={rgb, 255:red, 0; green, 0; blue, 0 }  ][line width=0.75]    (10.93,-3.29) .. controls (6.95,-1.4) and (3.31,-0.3) .. (0,0) .. controls (3.31,0.3) and (6.95,1.4) .. (10.93,3.29)   ;
\draw    (300,157) .. controls (292.12,178.67) and (281.33,181.91) .. (256.16,156.19) ;
\draw [shift={(255,155)}, rotate = 46.08] [color={rgb, 255:red, 0; green, 0; blue, 0 }  ][line width=0.75]    (10.93,-3.29) .. controls (6.95,-1.4) and (3.31,-0.3) .. (0,0) .. controls (3.31,0.3) and (6.95,1.4) .. (10.93,3.29)   ;
\draw    (155,158) .. controls (147.12,179.67) and (133.42,185.82) .. (108.16,160.19) ;
\draw [shift={(107,159)}, rotate = 46.08] [color={rgb, 255:red, 0; green, 0; blue, 0 }  ][line width=0.75]    (10.93,-3.29) .. controls (6.95,-1.4) and (3.31,-0.3) .. (0,0) .. controls (3.31,0.3) and (6.95,1.4) .. (10.93,3.29)   ;

\draw (90,127) node [anchor=north west][inner sep=0.75pt]  [color={rgb, 255:red, 255; green, 255; blue, 255 }  ,opacity=1 ] [align=left] {\fontfamily{ptm}\selectfont \textbf{\large{0}}};
\draw (197,98) node [anchor=north west][inner sep=0.75pt]   [align=left] {$1$};
\draw (272,98) node [anchor=north west][inner sep=0.75pt]   [align=left] {$1$};
\draw (199,159) node [anchor=north west][inner sep=0.75pt]   [align=left] {$1$};
\draw (277,159) node [anchor=north west][inner sep=0.75pt]   [align=left] {$1$};
\draw (129,159) node [anchor=north west][inner sep=0.75pt]   [align=left] {$1$};
\draw (163,125) node [anchor=north west][inner sep=0.75pt]   [align=left] {$ $};
\draw (160,125) node [anchor=north west][inner sep=0.75pt]   [align=left] {\fontfamily{ptm}\selectfont \textbf{\large{1}}};
\draw (232,125) node [anchor=north west][inner sep=0.75pt]   [align=left] {\fontfamily{ptm}\selectfont \textbf{\large{2}}};
\draw (306,125) node [anchor=north west][inner sep=0.75pt]   [align=left] {\fontfamily{ptm}\selectfont \textbf{\large{3}}};

\end{tikzpicture}
\caption{(color online) A continuous time random walker on a 4-site lattice with jump rate $1$, where the site $0$ is absorbing, the site $3$ is reflecting, and the site $2$ (shaded in blue) is where residence time is measured.}
\label{fig:fourabsorbing}
\end{figure}

The Brownian motion with an absorbing wall studied in \sref{sec:residencecontinuous} can be viewed as a continuous limit of a continuous-time random walk on a lattice with an absorbing site at $i=0$ and a reflecting wall at the rightmost site. The issue regarding the absorbing boundary condition can be understood already on a four-site problem, illustrated in \fref{fig:fourabsorbing}. 

The observable of interest is the residence time at the site $i=2$, defined by $Q=\sum_1^T\delta_{C_t,2}$. The corresponding scgf is the largest eigenvalue of the tilted  matrix 
\begin{equation}
    \begin{pmatrix}
    0 & 1 & 0 & 0 \\
    0  & -2  & 1 & 0 \\
    0  & 1  & p-2 & 1 \\
    0 & 0 & 1 & -1
    \end{pmatrix}.
\end{equation}
which is constructed following the discussion in the Appendix B of \cite{derrida2019large}.
A left eigenvector $(y_0, y_1,y_2,y_3)$ for the eigenvalue $\lambda$ follows the equation
\begin{equation}
    \begin{pmatrix}
        0 \\
        y_0-2y_1+y_2\\
        y_1-2y_2+y_3+py_2 \\
        y_2-y_3
    \end{pmatrix}=
    \lambda\begin{pmatrix}
        y_0 \\
        y_1\\
        y_2\\
        y_3
    \end{pmatrix},\label{eq:RW left eigenvalue}
\end{equation}
where the matrices have been transposed for the sake of legibility. We see from $\lambda y_0=0$, that a non-zero eigenvalue demands $y_0=0$, which in the continuous limit corresponds to the absorbing boundary condition in Sec \ref{sec:absorbingBC}. For the eigenvalue $\lambda=0$, the solution of \eqref{eq:RW left eigenvalue} does not impose vanishing boundary condition at the absorbing site. 

In a suitable continuous limit, both eigenvectors satisfy reflecting boundary condition at the rightmost end.

\subsection{Large deviation for the epidemiological process}
\label{sec:multievaporate_derivation}
In this section we present a derivation of \eqref{eq:ldf eva exact}. The probability of $Q$ in \eqref{eqn:evapobservable} with an initial $N$ number of people,
\begin{equation}
    P(Q)=\sum_{k=0}^{N}P(Q,k),
    \label{eqn:trajsumevap}
\end{equation}
where $P(Q,k)$ is the joint probability of $Q$ and that $k$ number of individuals survived till time $T$. 

\begin{figure}[h]
\centering
\begin{tikzpicture}[x=0.6pt,y=0.6pt,yscale=-1,xscale=1]

\draw [line width=3]    (152,128) -- (532,125) ;
\draw    (152,121) -- (152,136) ;
\draw    (531,117) -- (531,132) ;
\draw    (197,120) -- (197,135) ;
\draw    (240,120) -- (240,135) ;
\draw    (310,120) -- (310,135) ;
\draw    (460,118) -- (460,133) ;

\draw (526,133.4) node [anchor=north west][inner sep=0.75pt]    {$T$};
\draw (147,135.4) node [anchor=north west][inner sep=0.75pt]    {$0$};
\draw (187,135.4) node [anchor=north west][inner sep=0.75pt]    {$t_{1}$};
\draw (232,135.4) node [anchor=north west][inner sep=0.75pt]    {$t_{2}$};
\draw (300,135.4) node [anchor=north west][inner sep=0.75pt]    {$t_{3}$};
\draw (450,135.4) node [anchor=north west][inner sep=0.75pt]    {$t_{N-k}$};
\draw (368,135.4) node [anchor=north west][inner sep=0.75pt]  [font=\Large]  {$.\ .\ .$};

\draw (168,105.4) node [anchor=north west][inner sep=0.75pt]  {$N$};
\draw (197,105.4) node [anchor=north west][inner sep=0.75pt]  {$N-1$};
\draw (255,105.4) node [anchor=north west][inner sep=0.75pt]  {$N-2$};
\draw (484,105.4) node [anchor=north west][inner sep=0.75pt]  {$k$};

\end{tikzpicture}

\caption{A sequence of deaths at times $t_1,\ldots, t_{N-k}$, with the number of living individuals indicated above the time-line.}
\label{fig:seq_death}
\end{figure}
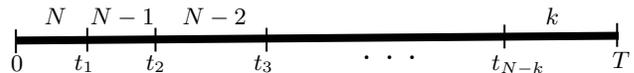

For evaluating this joint probability, consider a sequence of individual deaths at times $t_1, t_2, \ldots, t_{N-k}$ in an increasing order of time in the interval $[0,T]$ (see \fref{fig:seq_death}). 

In the interval $(t_i, t_{i+1})$, there are $(N - i)$ living persons, who contribute an amount $(N - i)(t_{i+1} - t_i)$ to the observable in \eqref{eqn:evapobservable}. The death rate in this interval is $(N - i)$, and as a result, the probability of all of them surviving in this interval and precisely one of them dying in the infinitesimal interval between $t_{i+1}$ and $t_{i+1} + dt_{i+1}$ is $\exp{-(N - i)^2(t_{i+1} - t_i)}(N - i)^2 dt_{i+1}$. Taking into account all intervals, we write
\begin{equation}
\begin{split}
P(Q,k)\sim &\int_{0}^{T}\mathbf{dt}\exp\left(-\sum_{i=0}^{N-k}(N-i)^2(t_{i+1}-t_{i})  \right) \\
    & \qquad \delta\left(Q - \sum_{i=0}^{N-k} (N-i)(t_{i+1}-t_{i}) \right), 
\end{split}\label{eqn:multi_eva_pdf}
\end{equation}
with $\mathbf{dt}=\prod_{i=1}^{N-k} dt_i$ subject to the constraint $t_{i}\le t_{i+1}$, $t_0=0$, and $t_{N-k+1}=T$, where we ignored algebraic pre-factors which are irrelevant for large deviation asymptotic.

For our analysis it is convenient to make a change of variables $t_i=T\, r_{N+1-i}$, and rewrite \eqref{eqn:multi_eva_pdf} as
\begin{equation}
\begin{split}
    P(q T,k)\sim &\int_{0}^{1}dr_{k+1}\cdots dr_{N}\exp\left(-T\sum_{i=k}^{ N}i^2(r_{i}-r_{i+1})  \right) \\
    & \qquad\delta\left(\sum_{i=k}^{N}  i(r_{i}-r_{i+1})-q \right), 
\end{split}\label{eqn:multi_eva_pdf_rescale}
\end{equation}
with $r_k=1$, $r_{N+1}=0$, and $r_{i}\ge r_{i+1}$.

Evidently, for large $T$, the integral in \eqref{eqn:multi_eva_pdf_rescale} is dominated by the minimum of the term in the exponential subjected to the constraint in the delta function. This gives the large deviation asymptotic $P(qT,k)\sim e^{-T \phi_k(q)}$ with
\begin{equation}
    \phi_k(q)=\min_{\{r_i\}}\sum_{i=k}^{ N}i^2(r_{i}-r_{i+1})
\label{eq:ldf variational 0}
\end{equation}
subjected to constraint $\sum_{i=k}^{N}  i(r_{i}-r_{i+1})=q$.

To solve this variational problem, we rewrite \eqref{eq:ldf variational 0} using $r_1=1$ and $r_{N+1}=0$, resulting in
\begin{equation}
    \phi(q)=\min_{\{r_i\}}\sum_{i=k+1}^{N}(2i-1)r_{i} +k^2\label{eq:ldf variational 1}
\end{equation}
with constraint $\sum_{i=k+1}^{N} r_{i}+k=q$. Considering that each term in the summation in \eqref{eq:ldf variational 1} is positive, the minimum is achieved by setting as many $r_i=0$ as possible. The integration domain $r_i \ge r_{i+1}$ imposes that, for a minimal solution, $r_i=0$ for all $i \ge i^\star$, as dictated by the value of $q$ in the constraint $\sum_{i=k+1}^{i^\star-1} r_i + k = q$.

Evidently, for $q < k$ and $q > N$, the constraint cannot be satisfied, implying $P(qT, k) = 0$ and, consequently, $\phi_k(q) \to \infty$. For $k \le q < k + 1$, the minimum solution corresponds to $i^\star = k + 2$ and $r_{k+1} = q - k$. Similarly, for $k + 1 \le q < k + 2$, the minimum corresponds to $i^\star = k + 3$, $r_{k+1} = 1$, $r_{k+2} = q - k - 1$, and $r_i = 0$ for the rest; and so on. This gives the solution of the variational problem
\begin{equation}
    \phi_k(q)=\{2(k+n)+1\}q-(k+n)(k+n+1)
\end{equation}
for $k+n\le q< k+n+1$ with $n=0,\cdots,N-k-1$.

Using this asymptotic of $P(Q,k)$ in \eqref{eqn:trajsumevap} yields the large deviation asymptotic \eqref{eqn:largedeviation} for $P(Q)$ with $\phi(q)=\min\{\phi_0(q), \ldots, \phi_N(q) \}$ leading to the expression \eqref{eq:ldf eva exact}.

\subsection{Large deviation for the mortal Brownians}
\label{sec:stickyderivation}

Here, we present a derivation for the piece-wise large deviation function discussed in \sref{sec:multisticky}. For this problem, the joint probability (analogous to \eqref{eqn:multi_eva_pdf}) is
\begin{equation*}
\begin{split}
P(Q,k)\sim\int_{0}^{T}\mathbf{dt}&\exp\left(-\sum_{i=0}^{N-k}(N-i)^2(t_{i+1}-t_{i})  \right) \\
  \int\limits_{-\infty}^{\infty}\mathbf{dx} & \exp  \left(-\sum_{i=1}^{ N}\frac{x_i^2}{4t_i}   \right) \delta\left(Q - \sum_{i=0}^{N} x_i \right)
\end{split}\label{eqn:multi_eva_pdf_2}
\end{equation*}
with $t_i=T$ for $i>N-k$, $\mathbf{dt}=dt_1\ldots dt_N$ and $\mathbf{dx}\equiv d x_1\ldots dx_{N}$, where we have used the Gaussian propagator for Brownians of unit diffusivity. From this point onward, the analysis follows a similar approach to that discussed in Appendix~\ref{sec:multievaporate_derivation}. For brevity, we directly present the final expression of the large deviation function.

The probability of $Q$ for large time $T$ follows the asymptotics \eqref{eqn:largedeviation} with a piece-wise ldf
\begin{equation}
    \phi(q)= \begin{dcases} 
    \frac{q^2}{4k}+k^2  & \textrm{for } \ell_{k-1} \leq |q| \leq \frac{k\ell_k}{k+1}\cr
    |q|\sqrt{2k+1}-k(k+1)& \textrm{for } \frac{k\ell_k}{k+1}\leq |q| \leq \ell_k,
    \end{dcases}
    \label{eqn:multistickyratefn}
\end{equation}
where $k=0,1,\ldots,N$, $\ell_k=2(k+1)\sqrt{2k+1}$ for all $k<N$, and $\ell_N$ infinite.

\subsection{Expressions of quantities defined in \sref{sec:localabsorbing}}\label{app:expressions}

The solutions $S_s$ and $F_s$ for the local time observable are obtained by solving (\ref{eq:eq for S}) and (\ref{eq:eq for F}), and are given by \eqref{eq:sticky local S and F}, with the explicit expressions for
\begin{widetext}
\begin{equation}
n_1(u,p,x_0)=\begin{dcases}
    \frac{1}{u^2}\left(d(u,p)-2 p \sinh \left(\frac{a u}{2}\right) \cosh \left(u (L-a)\right) \cosh \left(\frac{1}{2} u (a-2
   x_0)\right)- u\cosh \left(u (x_0-L)\right)\right) & \text{for }x_0 \leq a, \\ 
    \frac{1}{u^2}\left(d(u,p)-\left(p \sinh \left(a u\right)+u\right) \cosh \left(u (x_0-L)\right)\right) & \text{for } x_0> a,
    \end{dcases}
    \end{equation}
    and
\begin{equation}
n_2(u,p,x_0)=\begin{dcases}
    u\cosh \left(u (x_0-L)\right)-p \cosh \left(u (a-L)\right) \sinh \left(u (x_0-a)\right) & \text{for } x_0 \leq a, \\
   u\cosh \left(u (x_0-L)\right) & \text{for } x_0> a,
    \end{dcases}
\end{equation}
\end{widetext}
with $d(u,p)$ defined in \eqref{eq:abs denominator d}.



\nocite{*}

\bibliography{references}

\end{document}